\documentclass[useAMS,usenatbib, usegraphicx]{mn2e}
\usepackage{amssymb}

\newcommand{\rmd}{{\rm{d}}}
\newcommand{\rme}{{\rm{e}}}
\newcommand{\rmi}{{\rm{i}}}

\newcommand{\oomega}{{\tilde\omega}}
\newcommand{\mmu}{{\tilde\mu}}
\newcommand{\mass}{\mu}

\newcommand{\cmnt}[1]{}

\newcommand{\smfrac}[2]{{\!\!\!\begin{array}c\frac{#1}{#2}\end{array}\!\!\!}}

\begin{document}

\title[Lindblad resonances I]{Lindblad resonance torques in relativistic discs: I. Basic equations}

\author[Hirata]
 {Christopher M. Hirata
\\Caltech M/C 350-17, Pasadena, California 91125, USA}

\date{22 February 2011}

\pagerange{\pageref{firstpage}--\pageref{lastpage}} \pubyear{2010}

\maketitle

\label{firstpage}

\begin{abstract}
Lindblad resonances have been suggested as an important mechanism for angular momentum transport and heating in discs in binary black hole systems.  We present the basic equations for the torque and heating rate for relativistic thin discs subjected to a perturbation.  The Lindblad resonance torque is written explicitly in terms of metric perturbations for an equatorial disc in a general axisymmetric, time-stationary spacetime with a plane of symmetry.  We show that the resulting torque formula is gauge-invariant.  Computations for the Schwarzschild and Kerr spacetimes are presented in the companion paper.
\end{abstract}


\section{Introduction}

The past several years have seen a surge in interest related to the electromagnetic signatures of merging black holes.  Such a signature would have to come not from the black holes themselves, but from the gas that surrounds them.  Heating of this gas and consequent emission of electromagnetic radiation has been discussed both in the context of the inspiral phase \citep{2009arXiv0906.0825C}, the coalescence \citep{2008PhRvL.101d1101K}, and in the post-merger phase as the mass loss and kick of the final black hole modify the orbits of the gas particles \citep{2007APS..APR.S1010B, 2008ApJ...682..758S, 2008ApJ...684..835S, 2010PhRvD..81d4004A, 2010MNRAS.401.2021R}.

It is often suggested that torques arising from Lindblad resonances\footnote{Other resonances may also be relevant, e.g. it has been suggested that there could be matter at the L4 and L5 Lagrange points of binary black holes \citep{2010arXiv1006.0182S}, but they require a fundamentally different treatment and will not be investigated here.} play a key role in redistributing gas in the inspiral phase \citep{2002ApJ...567L...9A, 2005ApJ...622L..93M, 2008ApJ...672...83M, 2009arXiv0906.0825C} and controlling the surface density profile and heating rate of the gas disc.  These torques act by exciting density perturbations at the location of either inner or outer Lindblad resonances (ILRs or OLRs), at which the synodic period (i.e. the time between successive passages of the secondary black hole and a disc particle) is an integer multiple of the period of radial epicyclic oscillations in the disc.  In some scenarios, the resonant torques operate in the nonrelativistic Newtonian regime, which has a long history of study in the context of galactic discs, planetary rings, and circumstellar discs \citep[e.g.][]{1972MNRAS.157....1L, 1978ApJ...222..850G, 1979ApJ...233..857G, 1980ApJ...241..425G, 1979MNRAS.186..799L}.  However, in others -- particularly the cases of inner discs \citep{2009arXiv0906.0825C} -- Lindblad resonant torques are used all the way in to radii of a few$\times 10M$.  In these cases, it is desirable to revisit the Lindblad resonances in a fully relativistic context.  This is especially true since pericentre precession introduces an additional ILR (the $m=1$ or 0:1 ILR) that has no analogue in the Newtonian-Keplerian problem.  The principal purpose of this paper and its companion is to provide a relativistic treatment of the Lindblad torques, including computation of the torque formula in black hole spacetimes (Schwarzschild or Kerr), in the extreme mass ratio limit.

This paper and its companion are not concerned with a full analysis of any one scenario for the generation of an electromagnetic counterpart to a black hole merger, although they are most relevant to the proposal of \citet{2009arXiv0906.0825C}.  Rather, our motivation is to establish the relativistic Lindblad torque formula so that it can be used to establish the role (or lack thereof) of Lindblad torques in future work.  In this paper (``Paper I''), we develop the general formalism for Lindblad torques in thin discs orbiting in the equatorial planes of axisymmetric, time-independent spacetimes with a plane of symmetry, and with weak perturbations of general form respecting the equatorial reflection symmetry.  This covers the case of a binary Schwarzschild black hole with an extreme mass ratio ($q=M_2/M_1\ll 1$) and a gas disc orbiting in the same plane.  It also covers the Kerr case if the primary hole's spin is aligned with the orbital angular momentum of the binary and disc (which may or may not be the physical case; here it is a simplifying assumption that we may wish to remove in future work).  We work out the torque formula in terms of the background metric and its perturbation $h_{\alpha\beta}$ and establish generic features such as gauge invariance.  The companion (``Paper II'') focuses on the specific cases of interest -- the Schwarzschild and Kerr metrics with a small perturber -- and describes the numerical evaluation of the resonant torque.

Our analysis considers the case of geometrically thin discs.  The alternative -- a geometrically thick disc, such as that in an advection-dominated accretion flow (ADAF) -- cannot be treated by the methods described here.  A more appropriate model for an extreme mass ratio binary where the secondary orbits within a thick disc was considered by \citet{2000ApJ...536..663N}.  We note however that one conceivable way to produce such a thick disc, even at initially high accretion rates as considered by \citet{2009arXiv0906.0825C}, would be for resonant heating to destroy the thin disc solution and result in a radiatively inefficient inner disc.  Assessment of this possibility requires us to be able to quantitatively compute the resonant torques.

We evaluate the torque here by assuming a particle disc, since previous works on Lindblad resonances have found that the specific dissipation mechanism (e.g. viscosity or propagation of spiral density waves as occurs in a hydrodynamic disc) does not affect the {\em total} torque at a resonance so long as the excitation of disc modes is localized near the resonance and in the linear regime \citep[e.g.][]{1987Icar...69..157M, 1998ApJ...504..983L, 2007MNRAS.374..131O}.  The underlying reason for this -- namely, that the vector eccentricity\footnote{The vector eccentricity is the eccentricity weighted by the direction of the pericentre, or $e\rme^{\rmi\varpi}$ in Keplerian elements.  In a non-Keplerian potential the longitude of pericentre precesses, but the vector eccentricities of particles at the same epoch may still be summed.} {\em integrated} over the resonance in each sector of the disc, which is both excited by the external perturbation and acted upon by the perturbation to yield the overall torque, is not changed but is simply redistributed by short-range interactions among disc particles -- is generic and we expect it to also hold in the relativistic case.  We also note that the modes of oscillation of relativistic discs have been investigated \citep[e.g.][]{1997ApJ...476..589P, 2001ApJ...548..335S, 2002ApJ...567.1043O}, however their excitation by perturbations to the spacetime have not yet been treated.

This paper is organized as follows.  Section~\ref{sec:bkg} lays out the assumed background spacetime and the motion of test particles in it.  Section~\ref{sec:T1} describes the behaviour of particles under a general perturbation to their Hamiltonian (gravitational or otherwise) and the resulting torque on an initially axisymmetric disc.  Section~\ref{sec:T2} re-expresses this torque in terms of the metric perturbation and demonstrates gauge invariance; it also gives a useful alternative expression for the torque in terms of the power delivered to a test particle on a slightly eccentric orbit.  Section~\ref{sec:K} shows that our expression reduces to the familiar expression for Lindblad torques in the familiar Newtonian-Keplerian case, i.e. in the spacetime of a point mass at radii $r/M\gg 1$.  Section~\ref{sec:heat} describes the disc heating at the resonance, and Section~\ref{sec:disc} concludes.

We use relativistic units where $G=c=1$.

\section{Background spacetime and particle trajectories}
\label{sec:bkg}

\subsection{The spacetime}

We consider the unperturbed problem of a disc orbiting in the equatorial plane of a black hole.  In the equatorial plane, the metric may be written as \citep[e.g.][]{1974ApJ...191..499P}
\begin{equation}
\rmd s^2 = -\rme^{2\nu} \rmd t^2 + \rme^{2\psi}(\rmd\phi-\oomega\rmd t)^2 + \rme^{2\mmu}\rmd r^2 + \rmd z^2,
\label{eq:metric}
\end{equation}
where $\nu$, $\psi$, $\oomega$, and $\mmu$ are functions of $r$; as $r\rightarrow\infty$, we have $\oomega,\mmu,\nu\rightarrow 0$ and $\psi\rightarrow\ln r$.  Note that this formulation is only sufficient for eccentricity resonances; if we were to consider inclination resonances, we would have to include the ${\cal O}(z^2)$ terms in the metric.

The contravariant components of this metric are:
\begin{eqnarray}
g^{tt} &=& -\rme^{-2\nu}
\nonumber \\
g^{t\phi}=g^{\phi t} &=& -\oomega \rme^{-2\nu}
\nonumber \\
g^{\phi\phi} &=& \rme^{-2\psi} - \oomega^2\rme^{-2\nu}
\nonumber \\
g^{rr} &=& \rme^{-2\mmu} {\rm~~and}
\nonumber \\
g^{zz} &=& 1.
\label{eq:contravariant}
\end{eqnarray}

Equation~(\ref{eq:metric}) has a residual gauge degree of freedom in the sense that we may freely reparameterize $r\rightarrow f(r)$.  We fix this by requiring $\rme^{\nu+\psi+\mmu}=r$.  This choice is easily verified to be valid for the Schwarzschild coordinate system in the case of a nonrotating black hole, and for the Boyer-Lindquist coordinate system in the case of a rotating black hole.

\subsection{Particle trajectories}

We utilize the Hamiltonian formulation of the equations of motion for a particle.  As is well-known, the action for a particle of mass $\mass$ is $S=-\mass\int \rmd\tau$, where $\tau$ is the proper time along the particle trajectory.  For our purposes, the fastest route to the torque formula is {\em not} to use the covariant representation of the action parameterized by the affine parameter, but rather to explicitly parameterize the particle's trajectory using the coordinate time $t$, which is always possible outside the outer horizon.  This method, which explicitly keeps only the 3 physical degrees of freedom, is best suited to a perturbation analysis.  The formulation of the problem is standard -- the Lagrangian in coordinate time is the basis of the exposition by \citet{IP60}, and Hamiltonianization of the coordinate time is a standard technique in post-Newtonian calculations \citep[e.g.][]{1973PThPh..50..492O} -- but will be explicitly given here since we will need to refer to it repeatedly throughout the calculation.

Defining $u^\mu$ to be the 4-velocity, i.e. the forward-directed tangent vector to the particle's trajectory with $u_\alpha u^\alpha=-1$, we see that $\rmd\tau = \rmd t/u^t$ and hence the Lagrangian is
\begin{equation}
L \equiv \frac{\rmd S}{\rmd t} = - \frac \mass{u^t}.
\label{eq:lagrangian}
\end{equation}
The degrees of freedom of the particle are its spatial coordinates $x^i(t)$; we note that $u^t$ depends on the spatial coordinates $x^i$ and time coordinate $t$, and on the 3 spatial velocities $\dot x^i=\rmd x^i/\rmd t$.  The conjugate momenta are $\pi_i\equiv\partial L/\partial \dot x^i$.

Noting that $\dot x^i = u^i/u^t$, we see that varying $g_{\alpha\beta}u^\alpha u^\beta=-1$ at fixed $x^\alpha$ gives $g_{\alpha\beta} u^\alpha \delta u^\beta = 0$, or $u_\alpha \delta u^\alpha=0$.  Therefore,
\begin{equation}
0 = u_t \delta u^t + u_i \delta (u^t\dot x^i) = (u_t+u_i \dot x^i)\delta u^t + u^t u_i\delta \dot x^i.
\end{equation}
This implies that
\begin{equation}
\frac{\partial u^t}{\partial \dot x^i} = -\frac{u^tu_i}{u_t + u_i\dot x^i}
= -\frac{u^tu_i}{u_t + u_iu^i/u^t};
\end{equation}
recalling that by normalization of the 4-velocity $u_iu^i = -1-u_tu^t$ shows this to be equal to $(u^t)^2u_i$.  Therefore the conjugate momentum associated with Eq.~(\ref{eq:lagrangian}) is
\begin{equation}
\pi_i = \frac{\partial L}{\partial\dot x^i} = \frac \mass{(u^t)^2} (u^t)^2u_i = \mass u_i\equiv p_i,
\end{equation}
where $p_i$ are the spatial components of the covariant physical 4-momentum.  From now on, we will simply write $p_i$ and drop the $\pi_i$ notation.

The Hamiltonian is given by
\begin{equation}
H = p_i\dot x^i-L = \mass u_i\frac{u^i}{u^t} + \frac \mass{u^t} = -\mass u_t = -p_t,
\label{eq:hamiltonian}
\end{equation}
where again we used the normalization of the 4-velocity, $u_iu^i = -1-u_tu^t$.  Thus the Hamiltonian for the particle's motion is simply the energy seen by an observer moving orthogonally to the hypersurface of constant $t$.  From a dynamical perspective, the Hamiltonian should be thought of as depending on $t$, $x^i$, and $p_i$; the formula for $p_t$ is the mass-shell relation\footnote{The requirement that the particle travel forward in time implies that we use the negative branch of the square root.} (derived from normalization of ${\bmath u}$):
\begin{equation}
H(t,x^i,p_i) = \frac{g^{ti}p_i - \sqrt{(g^{ti}p_i)^2 - g^{tt}g^{ij}p_ip_j - \mass^2g^{tt}}}{g^{tt}}.
\label{eq:H-explicit}
\end{equation}

\subsection{Nearly circular, equatorial orbits}

We now consider the nearly circular orbits in the background space-time.  We restrict ourselves to equatorial orbits with $z=p_z=0$.

\subsubsection{Form of the Hamiltonian}

A circular orbit is a solution for which $\dot r=0$ or (equivalently) $p_r=0$.  We will be considering nearly circular orbits, i.e. we will expand the Hamiltonian to order $(p_r)^2$.  From Eqs.~(\ref{eq:contravariant}) and (\ref{eq:H-explicit}), we find that in general
\begin{equation}
H = \oomega p_\phi + \rme^\nu\sqrt{\mass^2 + \rme^{-2\mmu}(p_r)^2 + \rme^{-2\psi}(p_\phi)^2}.
\end{equation}
We now consider linear perturbations around a reference circular orbit.  To do this, we first expand to second order in $p_r$:
\begin{eqnarray}
H(p_\phi,r,p_r) &=& H(p_\phi,r,0) + \frac{\rme^{\nu-2\mmu}}{\sqrt{\mass^2 + \rme^{-2\psi}(p_\phi)^2}} \frac{(p_r)^2}2
\nonumber \\ && + {\cal O}[(p_r)^4],
\label{eq:X1}
\end{eqnarray}
where
\begin{equation}
H(p_\phi,r,0) = \oomega p_\phi + \rme^\nu\sqrt{\mass^2 + \rme^{-2\psi}(p_\phi)^2}.
\label{eq:X2}
\end{equation}
For a given value of $p_\phi$ one can find the minimum of $H(p_\phi,r,0)$ with respect to $r$, which (since $\partial H/\partial r=0$) corresponds to a circular orbit.\footnote{We consider only the stable solutions; maxima of $H$ or values of $p_\phi$ for which there is no circular orbit solution are not of interest here.}  We can expand around any such minimum (with $p_\phi=P_\phi$ and $r=R$) by writing
\begin{equation}
\Delta p_\phi = p_\phi - P_\phi {\rm ~~and~~} \Delta r = r-R.
\end{equation}
The transformation from $(r,\phi,p_r,p_\phi)$ to $(\Delta r, \phi, p_r,\Delta p_\phi)$ is a simple translation and hence is canonical.
Then $H(p_\phi,r,p_r)$ can then be Taylor-expanded around $(P_\phi,R,0)$:
\begin{equation}
H(\Delta p_\phi,\Delta r,p_r) = \!\! \sum_{\beta_1,\beta_2,\beta_3\ge 0} \!\! \frac{C_{\beta_1\beta_2\beta_3}
(\Delta p_\phi)^{\beta_1} \Delta r^{\beta_2} (p_r)^{\beta_3}
}{\beta_1!\beta_2!\beta_3!},
\end{equation}
where $C_{\beta_1\beta_2\beta_3}$ are the expansion coefficients. In order to study small perturbations of the orbits, we need to keep terms up to second order, i.e. $\beta_1+\beta_2+\beta_3\le 2$, and we drop those whose coefficients vanish. This leaves us with
\begin{eqnarray}
H(\Delta p_\phi,\Delta r,p_r) &=& C_{000} + C_{100}\Delta p_\phi + \frac12C_{200}(\Delta p_\phi)^2
\nonumber \\
&& + \frac12C_{020}\Delta r^2 + \frac12C_{002}(p_r)^2
\nonumber \\
&& + C_{110}\Delta p_\phi \Delta r.
\label{eq:Taylor}
\end{eqnarray}

\subsubsection{Relation of the coefficients to specific energy and angular momentum}
\label{ss:relations}

Some of the Taylor expansion coefficients in Eq.~(\ref{eq:Taylor}) have a straightforward interpretation, and all are calculable in terms of metric coefficients and the specific energy and angular momentum.  We denote the specific energy ($H/\mass$) and specific angular momentum ($p_\phi/\mass$) associated with a circular orbit of radius $r$ by ${\cal E}(r)$ and ${\cal L}(r)$.  We may also define ${\bmath w}$ to be the 4-velocity associated with the circular orbit.  Its covariant components are $w_t=-{\cal E}(R)$, $w_\phi={\cal L}(R)$, and $w_r=w_z=0$.  Using the inverse metric, the contravariant components are
\begin{equation}
w^t = \rme^{-2\nu}({\cal E}-\oomega{\cal L}),
\label{eq:wt}
\end{equation}
\begin{equation}
w^\phi = \oomega\rme^{-2\nu}({\cal E}-\oomega{\cal L}) + \rme^{-2\psi}{\cal L},
\label{eq:wphi}
\end{equation}
and $w^r=w^z=0$.

By definition,
\begin{equation}
C_{000} = \mass{\cal E}(R).
\end{equation}
If we consider a sequence of circular orbits parameterized by $r$, we may take the total derivatives of the Hamiltonian with respect to $r$ [i.e. derivatives in which $p_\phi=\mass{\cal L}(r)$ varies as we take the derivative]:
\begin{equation}
\frac{\rmd}{\rmd r}[\mass{\cal E}(r)] = \frac{\partial H}{\partial r} + \frac{\rmd [\mass{\cal L}(r)]}{\rmd r} \frac{\partial H}{\partial p_\phi},
\end{equation}
which since $\partial H/\partial r=0$ simplifies to
\begin{equation}
{\cal E}'(r) = {\cal L}'(r)\, \frac{\partial H}{\partial p_\phi}
\label{eq:temp100}
\end{equation}
or
\begin{equation}
C_{100} = \frac{{\cal E}'(R)}{{\cal L}'(R)} \equiv \Omega(R).
\end{equation}
We note that for a circular orbit, $\dot\phi = \partial H/\partial p_\phi=C_{100}=\Omega(R)$, so $\Omega(R)$ can be interpreted as the
angular frequency of the orbit as seen by a distant observer.  For the circular orbit, we also see that trivially
\begin{equation}
\Omega = \frac{w^\phi}{w^t}.
\label{eq:Omegafrac}
\end{equation}

Taking yet another total derivative of Eq.~(\ref{eq:temp100}) gives
\begin{eqnarray}
{\cal E}''(r) &=& {\cal L}''(r)\, \frac{\partial H}{\partial p_\phi}
\nonumber \\ &&
  + {\cal L}'(r) \left\{ \frac{\partial^2 H}{\partial r \partial p_\phi}
  + \frac{\rmd [\mass{\cal L}(r)]}{\rmd r} \frac{\partial^2 H}{\partial (p_\phi)^2}
  \right\},
\end{eqnarray}
or at $r=R$:
\begin{equation}
{\cal E}''(R) = {\cal L}''(R)\frac{{\cal E}'(R)}{{\cal L}'(R)} + {\cal L}'(R) \,[ C_{110} + \mass C_{200}{\cal L}'(R) ].
\end{equation}
Using the quotient rule, this can be expressed in terms of $\Omega(R)$:
\begin{equation}
\Omega'(R) = C_{110} + \mass C_{200}{\cal L}'(R).
\label{eq:X3}
\end{equation}

Similarly, taking the total derivative of the relation $\partial H/\partial r=0$ gives
\begin{equation}
\frac{\partial^2H}{\partial r^2} + \frac{\rmd [\mass{\cal L}(r)]}{\rmd r} \frac{\partial^2H}{\partial r\partial p_\phi} = 0.
\end{equation}
Evaluated at $R$, this simplifies to
\begin{equation}
C_{020} + \mass{\cal L}'(R) C_{110} = 0.
\label{eq:X4}
\end{equation}

A relation for $C_{002}$ can be obtained from Eq.~(\ref{eq:X1}):
\begin{equation}
C_{002} = \frac{\rme^{\nu-2\mmu}}{\sqrt{\mass^2 + \rme^{-2\psi}(p_\phi)^2}}.
\end{equation}
The value of the square root is obtainable from Eq.~(\ref{eq:X2}), giving
$C_{002} = \rme^{\nu-2\mmu}/[\rme^{-\nu}(H-\oomega p_\phi)]$, or
\begin{equation}
C_{002} = \frac{\mass^{-1}\rme^{2\nu-2\mmu}}{{\cal E}(R) - \oomega{\cal L}(R)} = \frac{\rme^{-2\mmu}}{\mass w^t}.
\label{eq:C002}
\end{equation}

Finally, we note that $C_{200}$ can be obtained by directly taking the second partial derivative of Eq.~(\ref{eq:X2}); noting that $\oomega$, $\nu$, and $\psi$ do not depend on $p_\phi$, we find
\begin{equation}
\left.\frac{\partial^2 H}{\partial(p_\phi)^2}\right|_{p_r=0} = \frac{\mass^2\rme^{\nu-2\psi}}{[\mu^2 + \rme^{-2\psi}(p_\phi)^2]^{3/2}},
\end{equation}
or at $r=R$:
\begin{equation}
C_{200} = \frac{\mass^{-1}\rme^{\nu-2\psi}}{[1 + \rme^{-2\psi}{\cal L}^2(R)]^{3/2}}.
\label{eq:X2.5}
\end{equation}
Further simplifcation is possible if we apply Eq.~(\ref{eq:X2}) to a circular orbit, yielding
\begin{equation}
{\cal E} = \oomega{\cal L} + \rme^\nu\sqrt{1 + \rme^{-2\psi}{\cal L}^2};
\end{equation}
since ${\cal E}-\oomega{\cal L} = \rme^{2\nu}w^t$, we conclude that
\begin{equation}
1 + \rme^{-2\psi}{\cal L}^2 = \rme^{2\nu}(w^t)^2.
\end{equation}
Substituting these results into Eq.~(\ref{eq:X2.5}) gives
\begin{equation}
C_{200} = \mass^{-1} \rme^{-2\nu-2\psi} (w^t)^{-3}.
\end{equation}

Combining with Eqs.~(\ref{eq:X3}) and (\ref{eq:X4}) gives
\begin{equation}
C_{110} = \Omega'(R) - \rme^{-2\nu-2\psi} (w^t)^{-3}{\cal L}'(R).
\end{equation}
and
\begin{equation}
C_{020} = -\mass\Omega'(R) {\cal L}'(R) + \mass\rme^{-2\nu-2\psi} (w^t)^{-3}{\cal L}'{^2}(R).
\label{eq:C020}
\end{equation}

Explicit evaluation of these expressions is aided by a relation for $w^t$.  Using the normalization $g_{\alpha\beta}w^\alpha w^\beta=-1$ and $w^\phi=\Omega w^t$, we find
\begin{equation}
w^t = [ \rme^{2\nu} - \rme^{2\psi}(\Omega-\oomega)^2 ]^{-1/2}.
\label{eq:wupt}
\end{equation}

This completes the description of the $C$ coefficients in terms of the commonly tabulated functions ${\cal E}(R)$, ${\cal L}(R)$, and $\Omega(R)$.  It is convenient also to define the epicyclic frequency
\begin{equation}
\kappa(R) \equiv \sqrt{C_{020}C_{002}};
\label{eq:kappaR}
\end{equation}
it is easy to see that if $\Delta p_\phi=0$, Eq.~(\ref{eq:Taylor}) guarantees that $\kappa(R)$ is the frequency of radial oscillations as measured by the coordinate time $t$.  We also define the specific epicyclic impedance
\begin{equation}
\mass{\cal Z}(R) \equiv \sqrt{\frac{C_{020}}{C_{002}}}.
\label{eq:ZR}
\end{equation}

\section{Relativistic resonant torque formula: formal solution}
\label{sec:T1}

\subsection{Perturbation Hamiltonian}

Our next concern is the canonical treatment of a perturbing body.  We separate the perturbation Hamiltonian into a perturbed and an unperturbed piece:
\begin{equation}
H(t,x^i,p_i) = H_0(t,x^i,p_i) + H_1(t,x^i,p_i).
\end{equation}
In Newtonian theory, the perturbing Hamiltonian $H_1$ is simply the gravitational potential of the perturbing body (plus an ``indirect term'' in formulations that do not use an inertial reference frame).  In GR, there is a perturbation to the metric:
\begin{equation}
g_{\alpha\beta} = g^{(0)}_{\alpha\beta} + h_{\alpha\beta} {\rm ~~~or~~~}
g^{\alpha\beta} = g^{(0)\alpha\beta} - h^{\alpha\beta},
\end{equation}
and $H_1$ is then the variation of Eq.~(\ref{eq:hamiltonian}) at fixed $p_i$,
\begin{equation}
H_1 = h^{\alpha\beta} \left.\frac{\partial p_t}{\partial g^{\alpha\beta}}\right|_{p_i}.
\label{eq:H1-1}
\end{equation}
The latter can be obtained by varying the mass-shell relation, $g^{\alpha\beta} p_\alpha p_\beta = -\mass^2$:
\begin{equation}
\delta g^{\alpha\beta}p_\alpha p_\beta + 2g^{\alpha\beta} p_\alpha \delta p_\beta=0.
\end{equation}
Since Eq.~(\ref{eq:H1-1}) is defined at fixed $p_i$, the last term may be restricted to $\beta=t$, and:
\begin{equation}
\delta p_t = -\frac1{2p^t} \delta g^{\alpha\beta}p_\alpha p_\beta = \frac{h^{\alpha\beta}p_\alpha p_\beta}{2p^t},
\end{equation}
where we have used the rule that the variation of the contravariant metric is $-g^{\alpha\kappa}g^{\beta\lambda}\delta g_{\kappa\lambda}$, i.e. the negative of $h$ with its indices raised.  Thus Eq.~(\ref{eq:H1-1}) simplifies to
\begin{equation}
H_1 = \frac{h^{\alpha\beta}p_\alpha p_\beta}{2p^t}.
\label{eq:H1}
\end{equation}

When doing perturbation theory, it is most convenient to do the explicit 3+1 expansion of the numerator and recall that
\begin{equation}
p^t = g^{tt} p_t + g^{ti} p_i = -g^{tt}H + g^{ti} p_i,
\end{equation}
so
\begin{equation}
H_1 = - \frac{h^{tt}H^2 + 2h^{ti}Hp_i + h^{ij}p_ip_j}{2(g^{tt}H - g^{ti} p_i)}.
\label{eq:H131}
\end{equation}
In first-order perturbation theory, it is permissible to replace $H$ with $H_0$ on the right-hand side of Eq.~(\ref{eq:H131}) since the latter already has one explicit power of $h$.

In the unperturbed case, the angular momentum $p_\phi$ and the energy $H$ are conserved.  In the perturbed case, these variables change in accordance with
\begin{equation}
\dot p_\phi = \{ p_\phi, H\}_{\rm P}= \{ p_\phi, H_1\}_{\rm P} = -\frac{\partial H_1}{\partial\phi}
\label{eq:lbracket}
\end{equation}
and
\begin{equation}
\dot H = \frac{\partial H}{\partial t} = \frac{\partial H_1}{\partial t},
\label{eq:ebracket}
\end{equation}
where $\{,\}_{\rm P}$ represents a Poisson bracket.
Since the perturbation arises from a secondary on a circular equatorial orbit, then the perturbation rotates at a pattern speed $\Omega_{\rm s}$ given by the orbit of the secondary hole, i.e. $H_1$ depends not on $t$ and $\phi$ individually but only on the combination $\phi-\Omega_{\rm s}t$.  This implies that the partial derivatives in Eqs.~(\ref{eq:lbracket}) and (\ref{eq:ebracket}) differ by a factor of $-\Omega_{\rm s}$, so
\begin{equation}
\dot H = -\Omega_{\rm s}\dot p_\phi.
\end{equation}

\subsection{Effect of perturbation on the disk}

We consider an ensemble of particles initially in circular orbit at radius $R$ with longitudes $\phi$ equally distributed in $\phi\in[0,2\pi)$.  The perturbation Hamiltonian is assumed to turn on at time $t_1$, and we wish to measure the torque on the disk of particles at some later time $t_2$.  The interval $t_2-t_1$ should be long compared with the orbital time $\Omega^{-1}$, but short compared with the libration time so that first-order perturbation theory for the positions of the particles is valid.  It is apparent that the torque $\langle T\rangle$ averaged over the ensemble of particles must be second-order in $h$ because if Eq.~(\ref{eq:lbracket}) is averaged over the unperturbed particle trajectories, we find
\begin{equation}
\langle T\rangle \equiv \langle \dot p_\phi \rangle = -\frac1{2\pi}\int_0^{2\pi} \frac{\partial H_1}{\partial\phi}\,\rmd\phi
= 0.
\end{equation}
In order to get a nonzero torque, we must compute the particle positions to first order in perturbation theory, and then apply Eq.~(\ref{eq:lbracket}).  This will lead to a result that is second order in $h$.

In what follows, we will construct the Green's function solution for the perturbations to the disk.  In order to evaluate the late-time torque, we will decompose the perturbation Hamiltonian into Fourier modes in the longitude direction,
\begin{equation}
H_1 = \sum_{m=-\infty}^\infty H_1^{(m)},
\end{equation}
where each mode has the dependence
\begin{equation}
H_1^{(m)} \propto \rme^{\rmi m (\phi - \Omega_{\rm s}t)}
\end{equation}
on longitude and time (the latter is required since the perturbation rotates with the orbit of the perturber).
Since the Hamiltonian is real, $H_1^{(m)\ast}=H_1^{(-m)}$.
The torque transfer from different values of $|m|$ can be considered separately.  This is because a first-order perturbation introduced by the $m$ component will have an $\rme^{\rmi m\phi(t_1)}$ longitude dependence, and hence can only produce an azimuthally averaged torque when acted on by the $m'$th Fourier mode of the perturbation if $m+m'=0$ or $m=-m'$.

Our final step in determining the first-order perturbation to the disk will be to integrate the Green's function over time, keeping only the resonant terms.

\subsubsection{Green's function solution for the perturbed disk}
\label{ss:Green}

We can compute the final position of a particle initially at $\phi_1\equiv\phi(t_1)$ via a Green's function method.  We consider first the effect of a $\delta$-function perturbation at time $t'$, i.e. we apply the perturbation
\begin{equation}
W(t) = H_1(t)\delta(t-t').
\end{equation}
Then to first order in perturbation theory, the perturbations to all variables can be written as an integral of the perturbation to that variable due to $W$ over the range $t_1<t'<t_2$.  Immediately prior to the application of $W$, the particle is at position
\begin{equation}
\phi(t'-\epsilon) = \phi_1 + \Omega(R)\,(t'-t_1).
\end{equation}
Immediately after the application of $W$, any phase-space coordinate $X$ undergoes a jump:
\begin{equation}
\diamondsuit X \equiv X(t'+\epsilon)-X(t'-\epsilon) = \{X,H_1(t')\}_{\rm P}.
\end{equation}
One key difference between this and Newtonian perturbation theory is that since $H_1$ depends on the momenta as well as the positions, the particle position can also undergo a jump.  These jumps are:
\begin{eqnarray}
\diamondsuit r &=& \frac{\partial H_1(t')}{\partial p_r},
\nonumber \\
\diamondsuit \phi &=& \frac{\partial H_1(t')}{\partial p_\phi},
\nonumber \\
\diamondsuit p_r &=& -\frac{\partial H_1(t')}{\partial r}, {\rm ~~and}
\nonumber \\
\diamondsuit p_\phi &=& -\frac{\partial H_1(t')}{\partial \phi}.
\end{eqnarray}

We then desire the final values of the positions and momenta.  These can be freely propagated from $t'+\epsilon$ using the unperturbed Hamiltonian, Eq.~(\ref{eq:Taylor}).  The angular momentum is the easiest since it is conserved
\begin{equation}
\Delta p_\phi(t_2) = \diamondsuit p_\phi = -\frac{\partial H_1(t')}{\partial\phi}.
\label{eq:delta-p-varphi}
\end{equation}
The radial degrees of freedom are more subtle.  They can be described by the Hamiltonian
\begin{equation}
H_0 = \frac12C_{020} \left( \Delta r + \frac{C_{110}}{C_{020}}\diamondsuit p_\phi \right)^2 + \frac12 C_{002}(p_r)^2 + {\rm const},
\end{equation}
which is identical to the Hamiltonian of a simple harmonic oscillator of effective spring constant $C_{020}$, effective mass $1/C_{002}$, and equilibrium position $-(C_{110}/C_{020})\diamondsuit p_\phi$. Under this Hamiltonian, the complex amplitude
\begin{equation}
Z \equiv \Delta r + \frac{C_{110}}{C_{020}}\diamondsuit p_\phi + \rmi \sqrt{\frac{C_{002}}{C_{020}}}\,p_r
\end{equation}
satisfies the equation of motion $\dot Z = -\rmi\kappa(R)Z$ and hence has a $\propto\rme^{-\rmi\kappa(R)\,t}$ dependence.  Its initial value is
\begin{equation}
Z(t'+\epsilon) = \frac{\partial H_1(t')}{\partial p_r} - \frac{C_{110}}{C_{020}}\frac{\partial H_1(t')}{\partial \phi}
 - \rmi \sqrt{\frac{C_{002}}{C_{020}}}\,\frac{\partial H_1(t')}{\partial r}.
\end{equation}
We may find $Z(t_2)$ by multiplying by $\rme^{-\rmi\kappa(R)\,(t_2-t')}$.  Taking the real and imaginary parts gives
\begin{eqnarray}
\Delta r(t_2) \!\!&=&\!\! \frac{C_{110}}{C_{020}} \frac{\partial H_1(t')}{\partial\phi}
\nonumber \\ &&\!\!
+ \left[ \frac{\partial H_1(t')}{\partial p_r} - \frac{C_{110}}{C_{020}}\frac{\partial H_1(t')}{\partial \phi}\right]\cos [\kappa(R)\,(t_2-t')]
\nonumber \\ &&\!\!
- \sqrt{\frac{C_{002}}{C_{020}}}\,\frac{\partial H_1(t')}{\partial r} \sin [\kappa(R)\,(t_2-t')]
\label{eq:delta-r}
\end{eqnarray}
and
\begin{eqnarray}
\Delta p_r(t_2) \!\!&=& \!\! -\frac{\partial H_1(t')}{\partial r}\cos  [\kappa(R)\,(t_2-t')]
\nonumber \\ &&\!\!\!
- \left[ \frac{\partial H_1(t')}{\partial p_r} - \frac{C_{110}}{C_{020}}\frac{\partial H_1(t')}{\partial \phi}\right]
\nonumber \\ &&\times
\sqrt{\frac{C_{020}}{C_{002}}} \sin [\kappa(R)\,(t_2-t')].
\label{eq:delta-p-r}
\end{eqnarray}
Finally, we may find the change in the longitude.  The change in its rate of advance can be found by varying $\partial H_0/\partial p_\phi$ using Eq.~(\ref{eq:Taylor}),
\begin{equation}
\Delta\dot\phi = C_{200}\Delta p_\phi + C_{110}\Delta r;
\end{equation}
and the change at time $t_2$ is found from
\begin{equation}
\Delta\phi(t_2) = \diamondsuit \phi + \int_{t'+\epsilon}^{t_2} \Delta\dot\phi(t_3) \rmd t_3.
\end{equation}
Using Eqs.~(\ref{eq:delta-p-varphi}) and (\ref{eq:delta-p-r}), we may perform the integral to get
\begin{eqnarray}
\Delta\phi(t_2) \!\!&=&\!\!
-\frac{\partial H_1(t')}{\partial p_\phi} - C_{200}(t_2-t')\frac{\partial H_1(t')}{\partial \phi}
\nonumber \\ && \!\!
+ C_{110}\Bigl\{\frac{C_{110}}{C_{020}} \frac{\partial H_1(t')}{\partial\phi} (t_2-t')
\nonumber \\ && \!\!
+ \left[ \frac{\partial H_1(t')}{\partial p_r} - \frac{C_{110}}{C_{020}}\frac{\partial H_1(t')}{\partial \phi}\right]\frac{\sin [\kappa(R)\,(t_2-t')]}{\kappa(R)}
\nonumber \\ && \!\!
- \sqrt{\frac{C_{002}}{C_{020}}}\,\frac{\partial H_1(t')}{\partial r} \frac{1-\cos [\kappa(R)\,(t_2-t')]}{\kappa(R)}\Bigr\}.
\label{eq:delta-varphi}
\end{eqnarray}

\subsubsection{Perturbed particle position for a particular Fourier mode of the perturbation}

At this point, we assume a particular Fourier mode $m$.  Then the perturbation Hamiltonians have a dependence $\rme^{\rmi m(\phi - \Omega_{\rm s}t)}$.  Since on the unperturbed trajectory, $\phi$ advances at a rate $\Omega(R)$, we may write
\begin{equation}
\frac{\partial H_1(t')}{\partial r} = \rme^{\rmi m[\Omega(R)-\Omega_{\rm s}](t'-t_2)}\frac{\partial H_1(t_2)}{\partial r},
\end{equation}
where the latter derivative is understood to be evaluated at the unperturbed longitude
\begin{equation}
\phi_2^{(0)} \equiv \phi_1 + \Omega(R)\,(t_2-t_1).
\end{equation}
[Note that the {\em actual} longitude is $\phi(t_2)=\phi_2^{(0)} + \Delta\phi(t_2)$.]  Inserting this dependence into Eqs.~(\ref{eq:delta-p-varphi}), (\ref{eq:delta-r}), (\ref{eq:delta-p-r}), and (\ref{eq:delta-varphi}) gives the following results for the $\delta$-function perturbation.  For the angular momentum,
\begin{equation}
\Delta p_\phi(t_2) = \diamondsuit p_\phi = - \rme^{\rmi m[\Omega(R)-\Omega_{\rm s}](t'-t_2)} \frac{\partial H_1(t_2)}{\partial\phi}.
\label{eq:delta-p-varphi2}
\end{equation}
For the radial position,
\begin{eqnarray}
\Delta r(t_2) \!\!&=&\!\! \frac{C_{110}}{C_{020}} \rme^{\rmi m[\Omega(R)-\Omega_{\rm s}](t'-t_2)} \frac{\partial H_1(t_2)}{\partial\phi}
\nonumber \\ &&\!\!
+ \left[ \frac{\partial H_1(t_2)}{\partial p_r} - \frac{C_{110}}{C_{020}}\frac{\partial H_1(t_2)}{\partial \phi}\right]
\nonumber \\ && \times \rme^{\rmi m[\Omega(R)-\Omega_{\rm s}](t'-t_2)}
\cos [\kappa(R)\,(t_2-t')]
\nonumber \\ &&\!\!
- \sqrt{\frac{C_{002}}{C_{020}}}\,\frac{\partial H_1(t_2)}{\partial r}
\nonumber \\ && \times
  \rme^{\rmi m[\Omega(R)-\Omega_{\rm s}](t'-t_2)} \sin [\kappa(R)\,(t_2-t')].
\label{eq:delta-r2}
\end{eqnarray}
For the radial momentum,
\begin{eqnarray}
\Delta p_r(t_2) \!\!&=& \!\! -\rme^{\rmi m[\Omega(R)-\Omega_{\rm s}](t'-t_2)} \frac{\partial H_1(t_2)}{\partial r}\cos  [\kappa(R)\,(t_2-t')]
\nonumber \\ &&\!\!\!
- \sqrt{\frac{C_{020}}{C_{002}}}
\left[ \frac{\partial H_1(t')}{\partial p_r} - \frac{C_{110}}{C_{020}}\frac{\partial H_1(t_2)}{\partial \phi}\right]
\nonumber \\ && \times
\rme^{\rmi m[\Omega(R)-\Omega_{\rm s}](t'-t_2)} \sin [\kappa(R)\,(t_2-t')].
\label{eq:delta-p-r2}
\end{eqnarray}
For the longitude,
\begin{eqnarray}
\Delta\phi(t_2) \!\!&=&\!\! \Bigl\{
-\frac{\partial H_1(t_2)}{\partial p_\phi} - C_{200}(t_2-t')\frac{\partial H_1(t_2)}{\partial \phi}
\nonumber \\ && \!\!
+ C_{110}\frac{C_{110}}{C_{020}} \frac{\partial H_1(t_2)}{\partial\phi} (t_2-t')
\nonumber \\ && \!\!
+ C_{110}\left[ \frac{\partial H_1(t_2)}{\partial p_r} - \frac{C_{110}}{C_{020}}\frac{\partial H_1(t_2)}{\partial \phi}\right]
\nonumber \\ && \times
\frac{\sin [\kappa(R)\,(t_2-t')]}{\kappa(R)}
\nonumber \\ && \!\!
- C_{110}\sqrt{\frac{C_{002}}{C_{020}}}\,\frac{\partial H_1(t_2)}{\partial r} \frac{1-\cos [\kappa(R)\,(t_2-t')]}{\kappa(R)}
\Bigr\}
\nonumber \\ && \times
\rme^{\rmi m[\Omega(R)-\Omega_{\rm s}](t'-t_2)}.
\label{eq:delta-varphi2}
\end{eqnarray}

\subsubsection{Integration of resonant terms}

We now integrate Eqs.~(\ref{eq:delta-p-varphi2}--\ref{eq:delta-varphi2}) over $\rmd t'$.  There are many terms, however most are of short period.  We therefore evaluate only the Lindblad resonant terms, i.e. those that satisfy the condition
\begin{equation}
m[\Omega(R)-\Omega_{\rm s}]\approx \pm \kappa(R).
\label{eq:LR}
\end{equation}
For positive $m$, the $+$ sign is appropriate for interior resonances and the $-$ sign for exterior; for negative $m$ this is reversed.\footnote{There are also corotation resonances where $\Omega(R)\approx\Omega_{\rm s}$, but we will not examine them here as the secondary hole actually orbits within the corotation resonance.}  It is convenient to write a resonant de-tuning function,
\begin{equation}
D(R) \equiv m[\Omega(R)-\Omega_{\rm s}] \mp \kappa(R).
\label{eq:D}
\end{equation}
Within this resonance condition, we may replace the time integral involving $\cos [\kappa(R)\,(t_2-t')]$:
\begin{eqnarray}
&&\!\!\!\! \int_{t_1}^{t_2} \rme^{\rmi m[\Omega(R)-\Omega_{\rm s}](t'-t_2)} \cos [\kappa(R)\,(t_2-t')] \rmd t'
\nonumber \\
&&\!\!\!\! \rightarrow \frac{1-\rme^{-\rmi D(R)\Delta t}}{2\rmi D(R)},
\end{eqnarray}
where $\Delta t\equiv t_2-t_1$.
A similar simplification occurs for the sine integral,
\begin{eqnarray}
&&\!\!\!\! \int_{t_1}^{t_2} \rme^{\rmi m[\Omega(R)-\Omega_{\rm s}](t'-t_2)} \sin [\kappa(R)\,(t_2-t')] \rmd t'
\nonumber \\
&&\!\!\!\! \rightarrow \mp\frac{1-\rme^{-\rmi D(R)\Delta t}}{2 D(R)}.
\end{eqnarray}
Only these factors have resonant denominators.  Integrating Eqs.~(\ref{eq:delta-p-varphi2}--\ref{eq:delta-varphi2}) gives no change in the angular momentum,
\begin{equation}
\Delta p_\phi(t_2)= 0;
\label{eq:int-pp}
\end{equation}
for the radial displacement,
\begin{eqnarray}
\Delta r(t_2) &=& \frac{1-\rme^{-\rmi D(R)\Delta t}}{2D(R)} \Bigl[
-\rmi\frac{\partial H_1^{(m)}(t_2)}{\partial p_r} 
\nonumber \\ &&
+\rmi \frac{C_{110}}{C_{020}}\frac{\partial H_1^{(m)}(t_2)}{\partial \phi}
\nonumber \\ &&
\pm \sqrt{\frac{C_{002}}{C_{020}}}\,\frac{\partial H_1^{(m)}(t_2)}{\partial r}
\Bigr];
\label{eq:int-r}
\end{eqnarray}
for the radial momentum,
\begin{eqnarray}
\Delta p_r(t_2) &=&  \frac{1-\rme^{-\rmi D(R)\Delta t}}{2D(R)} \Bigl[
\rmi\frac{\partial H_1^{(m)}(t_2)}{\partial r}
\nonumber \\ &&
\pm \sqrt{\frac{C_{020}}{C_{002}}} \frac{\partial H_1^{(m)}(t_2)}{\partial p_r}
\nonumber \\ &&
\mp \frac{C_{110}}{\sqrt{C_{020}C_{002}}} \frac{\partial H_1^{(m)}(t_2)}{\partial\phi}
\Bigr];
\label{eq:int-pr}
\end{eqnarray}
and for the longitude,
\begin{eqnarray}
\Delta\phi(t_2) &=& \frac{1-\rme^{-\rmi D(R)\Delta t}}{2D(R)} \frac{C_{110}}{\kappa(R)} \Bigl[
-\rmi\sqrt{\frac{C_{002}}{C_{020}}} \frac{\partial H_1^{(m)}(t_2)}{\partial r}
\nonumber \\ &&
\mp C_{110}
\frac{\partial H_1^{(m)}(t_2)}{\partial p_r} \pm \frac{C_{110}}{C_{020}}\frac{\partial H_1^{(m)}(t_2)}{\partial \phi}
\Bigr].
\label{eq:int-p}
\end{eqnarray}

\subsection{Net torque on the disk}

We are now ready to compute the torque exerted on the disk.  Since the torque on the unperturbed disk vanishes, we may compute the $\phi_2^{(0)}$-averaged torque on the first-order perturbed disk.  Recalling that only the $-m$ component of the perturbation gives an angle-averaged torque on the $m$ component of the perturbation, we find
\begin{eqnarray}
T^{(m)} &=& -\sum_{X\in\{r,\phi,p_r,p_\phi\}}\frac{\partial^2 H_1^{(-m)}(t_2)}{\partial\phi\partial X} \Delta X(t_2)
\nonumber \\
&=& \rmi m \sum_{X\in\{r,\phi,p_r,p_\phi\}}\frac{\partial H_1^{(m)\ast}}{\partial X}\Delta X(t_2).
\end{eqnarray}

Using Eqs.~(\ref{eq:int-pp}--\ref{eq:int-p}), we may evaluate this as:
\begin{eqnarray}
T^{(m)} &=& \rmi m \frac{1- \rme^{-\rmi D(R)\Delta t}}{2D(R)} \Bigl\{
\frac{\partial H_1^{(m)\ast}}{\partial r}
\Bigl[
-\rmi\frac{\partial H_1^{(m)}(t_2)}{\partial p_r}
\nonumber \\ &&
+\rmi \frac{C_{110}}{C_{020}}\frac{\partial H_1^{(m)}(t_2)}{\partial \phi}
\pm \sqrt{\frac{C_{002}}{C_{020}}}\,\frac{\partial H_1^{(m)}(t_2)}{\partial r}
\Bigr]
\nonumber \\ &&
+\frac{\partial H_1^{(m)\ast}}{\partial p_r}
 \Bigl[
\rmi\frac{\partial H_1^{(m)}(t_2)}{\partial r}
\pm \sqrt{\frac{C_{020}}{C_{002}}} \frac{\partial H_1^{(m)}(t_2)}{\partial p_r}
\nonumber \\ &&
\mp \frac{C_{110}}{\sqrt{C_{020}C_{002}}} \frac{\partial H_1^{(m)}(t_2)}{\partial\phi}
\Bigr]
\nonumber \\ &&
+\frac{\partial H_1^{(m)\ast}(t_2)}{\partial \phi} \frac{C_{110}}{\kappa(R)}
\Bigl[
-\rmi\sqrt{\frac{C_{002}}{C_{020}}} \frac{\partial H_1^{(m)}(t_2)}{\partial r}
\nonumber \\ &&
\mp \frac{\partial H_1^{(m)}(t_2)}{\partial p_r} \pm \frac{C_{110}}{C_{020}}\frac{\partial H_1^{(m)}(t_2)}{\partial \phi}
\Bigr]
\Bigr\}.
\label{eq:torque-1}
\end{eqnarray}
The quantity in braces $\{\}$ looks complicated, but if we substitute $\kappa(R)=\sqrt{C_{020}C_{002}}$ (c.f. Eq.~\ref{eq:kappaR}) it simplifies to
\begin{equation}
T^{(m)} = \pm\rmi m \frac{1- \rme^{-\rmi D(R)\Delta t}}{2D(R)}  \sqrt{\frac{C_{020}}{C_{002}}} |{\cal S}^{(m)}(t_2)|^2,
\label{eq:TM}
\end{equation}
where the interaction amplitude is
\begin{eqnarray}
{\cal S}^{(m)} &=& m\frac{C_{110}}{C_{020}}H_1^{(m)} \mp \sqrt{\frac{C_{002}}{C_{020}}} \frac{\partial H_1^{(m)}}{\partial r}
  + \rmi \frac{\partial H_1^{(m)}}{\partial p_r}
  \nonumber \\
&=& -\frac{m H_1^{(m)}}{\mass{\cal L}'(R)} \mp \frac1{\mass{\cal Z}(R)}\frac{\partial H_1^{(m)}}{\partial r}
  + \rmi \frac{\partial H_1^{(m)}}{\partial p_r}.
\label{eq:SM}
\end{eqnarray}
We note that while ${\cal S}^{(m)}$ is formally evaluated at time $t=t_2$, its time dependence is $\propto \rme^{-\rmi m\Omega_{\rm s}t}$ and hence its modulus $|{\cal S}^{(m)}(t)|$ is constant.  We can also relate ${\cal S}^{(m)}$ to ${\cal S}^{(-m)}$: since $H_1^{(-m)}=H_1^{(m)\ast}$, and since the sign of the resonant term [c.f. Eq.~(\ref{eq:D})] changes when we switch from $m$ to $-m$,
\begin{equation}
{\cal S}^{(-m)} = -{\cal S}^{(m)\ast}.
\end{equation}
We can then write the {\em total} torque arising from both the $m$ and $-m$ resonant terms as $T=T^{(m)}+T^{(-m)}$:
\begin{eqnarray}
T &=& \pm\rmi  \left[ m \frac{1- \rme^{-\rmi D(R)\Delta t}}{2D(R)} - (m \leftrightarrow -m) \right] 
\nonumber \\ && \times \sqrt{\frac{C_{020}}{C_{002}}} |{\cal S}^{(m)}|^2.
\end{eqnarray}
Since $D(R)$ flips sign between the $m$ and $-m$ resonances,
\begin{eqnarray}
T &=& \pm \rmi  \left[ m \frac{1- \rme^{-\rmi D(R)\Delta t}}{2D(R)} + 
m \frac{1- \rme^{\rmi D(R)\Delta t}}{-2D(R)} \right] 
\nonumber \\ && \times \sqrt{\frac{C_{020}}{C_{002}}} |{\cal S}^{(m)}|^2
\nonumber \\
&=& \mp m \frac{\sin [D(R)\,\Delta t]}{D(R)} \sqrt{\frac{C_{020}}{C_{002}}} |{\cal S}^{(m)}|^2.
\label{eq:T-total}
\end{eqnarray}

Equation~(\ref{eq:T-total}) has now separated into two pieces.  There is an $R$-dependent prefactor that contains the form of the resonance, and the factor ${\cal S}^{(m)}$ that encodes information on the normalization of the resonance and does not vary significantly across its width.  The first piece can be simplified by noting that it is dominated by regions with $|D(R)|\lesssim \Delta t^{-1}$.  Its integral is
\begin{equation}
\int  \frac{\sin [D(R)\,\Delta t]}{D(R)} 
 \rmd R = \frac{\pi }{|D'(R)|},
\end{equation}
so we approximate it near resonance as
\begin{equation}
\frac{\sin [D(R)\,\Delta t]}{D(R)} \rightarrow \frac{\pi }{|D'(R)|}\delta(R-R_{\rm r}),
\end{equation}
where $R_{\rm r}$ is the radius of exact resonance.  We thus find
\begin{equation}
T = \mp \frac{\pi m }{|D'(R)|} \mass{\cal Z}(R) |{\cal S}^{(m)}|^2 \delta(R-R_{\rm r}).
\label{eq:TX1}
\end{equation}

Often we want to know the torque density $\rmd T/\rmd r$.  For a thin disk with proper surface density $\Sigma$, i.e. whose 3-dimensional proper density is $\rho_0=\Sigma\delta(z)$, the rest mass per unit radius is
\begin{equation}
\frac{\rmd \mass}{\rmd r} = -\frac1{\Delta r}\int \rho_0 {\bmath w}\cdot{\bmath n} \rmd^3V,
\end{equation}
where $\rmd^3V$ represents the volume of a spacelike 3-surface spanning the range from $r$ to $r+\Delta r$ and ${\bmath n}$ is the unit forward-directed normal to this surface.  Taking the surface to be at constant $t$, the normal is $n_\alpha = (-\rme^\nu, 0,0,0)$ and the volume element is $\rmd^3V=\rme^{\psi+\mmu}\,\rmd r\,\rmd\phi\,\rmd z$.  This leads to the result
\begin{equation}
\frac{\rmd \mass}{\rmd r} = 2\pi rw^t\Sigma.
\end{equation}
It follows that
\begin{equation}
\frac{\rmd T}{\rmd r} = \mp \frac{2 \pi^2 m}{|D'| }rw^t{\cal Z}\Sigma |{\cal S}^{(m)}|^2 \delta(r-R_{\rm r}). 
\label{eq:dTdr}
\end{equation}

\section{Relativistic resonant torque formula: evaluation}
\label{sec:T2}

Having the formal solution for the torque (Eq.~\ref{eq:TX1}) is only part of the problem; we also need the resonant amplitude ${\cal S}^{(m)}$.  This section evaluates the amplitude and then shows that (within some restrictions) it is gauge-invariant.

\subsection{Evaluation of ${\cal S}^{(m)}$}

Here we require both the perturbation Hamiltonian, and its derivatives with respect to $r$ and $p_r$.  These are all to be evaluated at the unperturbed circular orbit using Eq.~(\ref{eq:H131}).

For $H_1$ itself, we see that since $H_0=\mass{\cal E}(R)$ and $p_\phi=\mass{\cal L}(R)$,
\begin{equation}
H_1 = \frac \mass2\rme^{2\nu} \frac{h^{tt}{\cal E}^2 - 2h^{t\phi}{\cal EL} + h^{\phi\phi}{\cal L}^2}{{\cal E}-\oomega{\cal L}}.
\end{equation}
For the partial derivatives with respect to the coordinates, we find that in general
\begin{eqnarray}
\frac{\partial H_1}{\partial x^k}
&=&
-\frac{h^{tt}{_{,k}}H_0^2 + 2h^{ti}{_{,k}}H_0p_i + h^{ij}{_{,k}}p_ip_j }{2(g^{tt}H - g^{ti} p_i)}
\nonumber \\
&&
+ \frac{2h^{tt}H_0(\partial H_0/\partial x^k)
  + 2h^{ti} (\partial H_0/\partial x^k) p_i}
  {2(g^{tt}H - g^{ti} p_i)}
\nonumber \\
&&
- H_1 \frac{g^{tt}(\partial H_0/\partial x^k) + g^{tt}{_{,k}}H_0 - g^{ti}{_{,k}}p_i}{g^{tt}H_0-g^{ti}p_i},
\end{eqnarray}
where the last term is associated with the derivative of the denominator in Eq.~(\ref{eq:H131}).
Since $\partial H_0/\partial r =0$ on a circular orbit, this implies
\begin{eqnarray}
\frac{\partial H_1}{\partial r} &=&
\frac \mass2\rme^{2\nu} \frac{h^{tt}{_{,r}}{\cal E}^2 - 2h^{t\phi}{_{,r}}{\cal EL} + h^{\phi\phi}{_{,r}}{\cal L}^2}{{\cal E}-\oomega{\cal L}}
\nonumber \\
&& + \left(2\nu_{,r}+\frac{\oomega_{,r}{\cal L}}{{\cal E}-\oomega{\cal L}} \right) H_1.
\end{eqnarray}

For the partial derivatives with respect to the momenta, the general expression is
\begin{eqnarray}
\frac{\partial H_1}{\partial p_k}
&=& \frac{(h^{tt}H_0+h^{ti}p_i)(\partial H_0/\partial p_k)  + h^{tk}H_0 + h^{ik}p_i}
  {g^{tt}H_0 - g^{ti}p_i}
\nonumber \\
&& - H_1 \frac{g^{tt}(\partial H_0/\partial p_k) - g^{tk}}{g^{tt}H_0-g^{ti}p_i}.
\end{eqnarray}
For the specific case of $p_r$, we note that at the circular orbit $\partial H_0/\partial p_r=0$ and $g^{tr}=0$, so
\begin{equation}
\frac{\partial H_1}{\partial p_r} = \frac{\rme^{2\nu}}{{\cal E}-\oomega{\cal L}}
(-h^{tr}{\cal E} + h^{r\phi}{\cal L}).
\end{equation}

We may now assemble the pieces to compute ${\cal S}^{(m)}$:
\begin{eqnarray}
{\cal S}^{(m)} &=& \frac{\rme^{2\nu}}{2({\cal E}-\oomega{\cal L})}\Bigl\{
\left[ -\frac m{{\cal L}'} \mp \frac1{\cal Z}\left(2\nu_{,r}+\frac{\oomega_{,r}{\cal L}}{{\cal E}-\oomega{\cal L}} \right) \right]
\nonumber \\ && \times
\left[h^{(m)tt}{\cal E}^2 - 2h^{(m)t\phi}{\cal EL} + h^{(m)\phi\phi}{\cal L}^2\right]
\nonumber \\ &&
\mp \frac1{\cal Z}\left[h^{(m)tt}{_{,r}}{\cal E}^2 - 2h^{(m)t\phi}{_{,r}}{\cal EL} + h^{(m)\phi\phi}{_{,r}}{\cal L}^2\right]
\nonumber \\ &&
-\rmi h^{(m)tr}{\cal E} + \rmi h^{(m)r\phi}{\cal L} \Bigr\}.
\label{eq:SM-explicit}
\end{eqnarray}
Note that this is independent of the particle mass $\mass$ and linear in the perturbation $h^{\alpha\beta}$.

It is possible to rewrite Eq.~(\ref{eq:SM-explicit}) in terms of the circular 4-velocity ${\bmath w}$.  Multiplying through by $w^t = \rme^{-2\nu}({\cal E}-\oomega{\cal L})$ gives
\begin{eqnarray}
w^t{\cal S}^{(m)} &=& \left[ -\frac m{{\cal L}'} \mp \frac1{\cal Z}\left(
2\nu_{,r} + \frac{\oomega_{,r}{\cal L}}{{\cal E}-\oomega{\cal L}}
\right)\right]
\frac {h^{(m)\alpha\beta}w_\alpha w_\beta }{2} 
\nonumber \\ && \mp \frac{h^{(m)\alpha\beta}{_{,r}}w_\alpha w_\beta}{2\cal Z}
+ \rmi h^{(m)r\alpha}w_\alpha.
\label{eq:FSM}
\end{eqnarray}
This form will be most useful in proving gauge invariance and in practical applications.

\subsection{Gauge invariance}

In general the perturbation $h^{\alpha\beta}$ could be expressed in many choices of gauge.  These differ by the relation
\begin{equation}
h_{\alpha\beta} \rightarrow h_{\alpha\beta} - \xi_{\alpha;\beta} - \xi_{\beta;\alpha}.
\end{equation}
Since Eq.~(\ref{eq:FSM}) is linear in $h_{\alpha\beta}$, the contributions to ${\cal S}^{(m)}$ from the pre-existing and gauge perturbations simply add, so to show invariance of the torque it is sufficient to prove that a pure gauge perturbation
\begin{equation}
h_{\alpha\beta} = -\xi_{\alpha;\beta} - \xi_{\beta;\alpha}
\label{eq:gauge}
\end{equation}
leads to zero resonant amplitude ${\cal S}^{(m)}$.  We restrict our attention to gauges that preserve the fundamental symmetries of the problem, i.e. that have reflection across the equatorial plane and have helical symmetry, where the $m$ Fourier component has an oscillatory time dependence $\propto \rme^{-\rmi \Omega_{\rm s}mt}$.  Without loss of generality, we may consider the Fourier modes one at a time, so we will consider the order $m$ Fourier mode below and avoid writing the superscript $^{(m)}$ explicitly.  Furthermore, it is easily seen that the $z$ coordinate is superfluous in computing Eq.~(\ref{eq:FSM}) in the equatorial plane, so we may restrict ourselves to the 2+1 dimensional equatorial slice of the spacetime.

While one could solve for  ${\cal S}^{(m)}$ for a pure gauge mode by explicit evaluation of Eq.~(\ref{eq:gauge}) followed by substitution into Eq.~(\ref{eq:FSM}), it is far easier to solve the problem by defining the combinations of metric perturbations and 4-velocities that appear in Eq.~(\ref{eq:FSM}):
\begin{eqnarray}
I_1 &\equiv& \frac12 h_{\alpha\beta} w^\alpha w^\beta,
\nonumber \\
I_2 &\equiv& \frac12h^{\alpha\beta}{_{,r}}w_\alpha w_\beta,
{\rm ~~and}
\nonumber \\
I_3 &\equiv& h^{r\alpha} w_\alpha,
\label{eq:I-exp}
\end{eqnarray}
and evaluating these in terms of $\bxi$ with the help of Lie derivatives.  We may then substitute into
\begin{equation}
w^t{\cal S} = \left[ -\frac m{{\cal L}'} \mp \frac1{\cal Z}\left(
2\nu_{,r} + \frac{\oomega_{,r}{\cal L}}{{\cal E}-\oomega{\cal L}}
\right)\right] I_1
\mp \frac{I_2}{\cal Z} + \rmi I_3,
\label{eq:FSM2}
\end{equation}
and then check whether the terms add to zero.

\subsubsection{Evaluation of $I_1$}

We begin by writing the equation for $h_{\alpha\beta}$ (Eq.~\ref{eq:gauge}) in the alternative form using the Lie derivative (e.g. Appendix C of \citealt{1984ucp..book.....W})
\begin{equation}
h_{\alpha\beta} = -\pounds_\bxi g_{\alpha\beta},
\label{eq:hab}
\end{equation}
or\footnote{The Lie derivative does {\em not} generally allow raising or lowering indices.  The raised-index relation arises by considering the inverse-metric formula $g^{\alpha\beta}g_{\beta\gamma}=\delta^\alpha_\gamma$.  Applying the product rule gives $(\pounds_\bxi g^{\alpha\beta})g_{\beta\gamma} + g^{\alpha\beta}\pounds_\bxi g_{\beta\gamma}=0$.  Substituting Eq.~(\ref{eq:hab}) and raising indices then gives $\pounds_\bxi g^{\alpha\beta} = h^{\alpha\beta}$.}
\begin{equation}
h^{\alpha\beta} = \pounds_\bxi g^{\alpha\beta} = \xi^\gamma g^{\alpha\beta}{_{,\gamma}}
  - g^{\gamma\beta} \xi^\alpha{_{,\gamma}} - g^{\alpha\gamma}\xi^\beta{_{,\gamma}}.
\label{eq:habinv}
\end{equation}

We are now in a position to compute the required term $h_{\alpha\beta} w^\alpha w^\beta$.  Noting that $g_{\alpha\beta}w^\alpha w^\beta=-1$, we take the Lie derivative $\pounds_\bxi$,
\begin{equation}
-h_{\alpha\beta} w^\alpha w^\beta + g_{\alpha\beta}(\pounds_\bxi w^\alpha) w^\beta
  + g_{\alpha\beta}w^\alpha (\pounds_\bxi w^\beta) = 0;
\end{equation}
using the symmetry in $\alpha$ and $\beta$ then gives
\begin{equation}
\frac12h_{\alpha\beta}w^\alpha w^\beta = g_{\alpha\beta}w^\alpha (\pounds_\bxi w^\beta) = {\bmath w}\cdot(\pounds_\bxi{\bmath w})
={\bmath w}\cdot[\bxi,{\bmath w}],
\label{eq:LX1}
\end{equation}
where [,] denotes a vector commutator.  We explicitly evaluate the $^t$ and $^\phi$ components of the commutator; recalling that $w^r=0$ and $w^t$ and $w^\phi$ depend only on $r$, we find
\begin{equation}
[\bxi,{\bmath w}]^t = \xi^r w^t{_{,r}} - w^t \dot \xi^t - w^\phi  \xi^t{_{,\phi}};
\end{equation}
using $\Omega=\dot\phi=w^\phi/w^t$ and the angular and time dependences of $\bxi$, we conclude that
\begin{equation}
[\bxi,{\bmath w}]^t = \xi^r w^t{_{,r}} + \rmi m w^t (\Omega_{\rm s} - \Omega)\xi^t.
\end{equation}
Similarly,
\begin{equation}
[\bxi,{\bmath w}]^\phi = \xi^r w^\phi{_{,r}} + \rmi m w^t (\Omega_{\rm s} - \Omega)\xi^\phi.
\end{equation}
Taking the dot product with ${\bmath w}$ gives
\begin{eqnarray}
{\bmath w}\cdot[\bxi,{\bmath w}] &=& \xi^r (w_tw^t{_{,r}} + w_\phi w^\phi{_{,r}})
\nonumber \\ &&
  + \rmi m(\Omega_{\rm s} - \Omega)w^t(w_t\xi^t + w_\phi\xi^\phi).
\end{eqnarray}
The first term evaluates to zero:
\begin{eqnarray}
w_tw^t{_{,r}} + w_\phi w^\phi{_{,r}} &=& -w^tw_t{_{,r}} - w^\phi w_\phi{_{,r}}
\nonumber \\ &=& -w^t(-{\cal E}') - (w^t\Omega){\cal L}' 
\nonumber \\
&=& w^t({\cal E}'-\Omega{\cal L}') = 0.
\end{eqnarray}
(The first equality can be shown by differentiating the relation $w_\alpha w^\alpha=-1$ with respect to $r$.)
The second is simplified using $w_t=-{\cal E}$ and $w_\phi={\cal L}$; thus we find that in general,
\begin{equation}
I_1 = \frac12h_{\alpha\beta}w^\alpha w^\beta = \rmi m(\Omega_{\rm s} - \Omega)w^t(-{\cal E}\xi^t + {\cal L}\xi^\phi).
\label{eq:LX2}
\end{equation}

\subsubsection{Evaluation of $I_2$}

We now turn our attention to $I_2$,
which appears in the second term in Eq.~(\ref{eq:FSM2}).  A reorganization gives
\begin{equation}
I_2 = \left(\frac12h^{\alpha\beta}w_\alpha w_\beta\right)_{,r} - h^{\alpha\beta}w_\alpha w_{\beta,r}.
\label{eq:LX3}
\end{equation}
Use of Eq.~(\ref{eq:LX2}) gives
\begin{eqnarray}
\left(\frac12h^{\alpha\beta}w_\alpha w_\beta\right)_{,r} \!\!\! &=& \!\!\!
\rmi m [-\Omega'w^t + (\Omega_{\rm s}-\Omega)w^t{_{,r}}]
\nonumber \\ && \!\!\!\times
(-{\cal E}\xi^t + {\cal L}\xi^\phi)
\nonumber \\ && \!\!\!
+ \rmi m(\Omega_{\rm s} - \Omega)w^t
\nonumber \\ && \!\!\!\times
(-{\cal E}'\xi^t + {\cal L}'\xi^\phi - {\cal E}\xi^t{_{,r}} + {\cal L}\xi^\phi{_{,r}}).
\label{eq:LX4}
\end{eqnarray} 

To complete the evaluation of $I_2$, we introduce the 1-form field 
\begin{equation}
s_\beta \equiv \pounds_{\partial/\partial r} w_\beta,
\end{equation}
whose components are $s_\beta = w_{\beta,r}$, or explicitly $s_t=-{\cal E}'$, $s_\phi={\cal L}'$, and $s_r=0$.  Then the last term in Eq.~(\ref{eq:LX3}) is $-h^{\alpha\beta} w_\alpha s_\beta$.  We can see that
\begin{equation}
g^{\alpha\beta}w_\alpha s_\beta = w^\beta s_\beta = -w^t{\cal E}'+w^\phi{\cal L}' = 0
\end{equation}
since $\Omega = {\cal E}'/{\cal L}' = w^\phi/w^t$.  Taking the Lie derivative $\pounds_\bxi$ gives
\begin{equation}
0 = h^{\alpha\beta}w_\alpha s_\beta + s^\alpha \pounds_\bxi w_\alpha + w^\alpha \pounds_\bxi s_\alpha.
\end{equation}
Rearranging and expanding the Lie derivatives gives
\begin{eqnarray}
- h^{\alpha\beta}w_\alpha s_\beta &=& s^\alpha \xi^\beta w_{\alpha,\beta} + s^\alpha w_\beta \xi^\beta{_{,\alpha}}
  + w^\alpha \xi^\beta s_{\alpha,\beta}
\nonumber \\ &&
  + w^\alpha s_\beta \xi^\beta{_{,\alpha}}.
\end{eqnarray}
Recalling that the terms containing $w_{\alpha,\beta}$ and $s_{\alpha,\beta}$ are only nonzero for $\beta=r$, that $w^\phi = \Omega w^t$, and the $\propto\rme^{\rmi m(\phi-\Omega_{\rm s}t)}$ dependence of the components of $\xi$, we reduce this to
\begin{eqnarray}
- h^{\alpha\beta}w_\alpha s_\beta &=& -s^t \xi^r {\cal E}' + s^\phi \xi^r {\cal L}'
\nonumber \\ &&
+ \rmi m (s^\phi - \Omega_{\rm s}s^t) (-{\cal E}\xi^t + {\cal L}\xi^\phi)
\nonumber \\ &&
- w^t \xi^r{\cal E}'' + w^\phi \xi^r {\cal L}''
\nonumber \\ &&
+ \rmi m (\Omega - \Omega_{\rm s})w^t (-{\cal E}'\xi^t + {\cal L}'\xi^\phi).
\end{eqnarray}
Combining this with Eq.~(\ref{eq:LX4}) gives
\begin{eqnarray}
I_2 &=&
\rmi m [-\Omega'w^t + (\Omega_{\rm s}-\Omega)w^t{_{,r}} + s^\phi - \Omega_{\rm s}s^t]
\nonumber \\ && \times
(-{\cal E}\xi^t + {\cal L}\xi^\phi)
\nonumber \\ && \!\!\!
+ \rmi m(\Omega_{\rm s} - \Omega)w^t
(- {\cal E}\xi^t{_{,r}} + {\cal L}\xi^\phi{_{,r}})
\nonumber \\ && \!\!\!
+ (
-s^t {\cal E}' + s^\phi  {\cal L}'
- w^t {\cal E}'' + w^\phi {\cal L}'' ) \xi^r.
\label{eq:I2-intermed}
\end{eqnarray}

Further simplification of this equation is possible using the contravariant components of ${\bmath s}$: raising indices gives
\begin{eqnarray}
s^t \!\! &=& \!\! \rme^{-2\nu}({\cal E}'-\oomega{\cal L}') {\rm ~~and}
\nonumber \\
s^\phi \!\! &=& \!\! \oomega \rme^{-2\nu}({\cal E}'-\oomega{\cal L}') + \rme^{-2\psi}{\cal L}'.
\label{eq:s-component}
\end{eqnarray}
From this we obtain
\begin{eqnarray}
-s^t {\cal E}' + s^\phi  {\cal L}' \!\! &=& \!\! -\rme^{-2\nu}({\cal E}'-\oomega{\cal L}')^2 + \rme^{-2\psi}{\cal L}'{^2}
\nonumber \\
\!\! &=& \!\!
{\cal L}'{^2} [ -\rme^{-2\nu}(\Omega-\oomega)^2 + \rme^{-2\psi} ]
\nonumber \\
\!\! &=& \!\!
\rme^{-2\nu-2\psi}(w^t)^{-2}{\cal L}'{^2},
\label{eq:slemma}
\end{eqnarray}
where the second line used $\Omega={\cal E}'/{\cal L}'$ and the third line used Eq.~(\ref{eq:wupt}). However, we also see that:
\begin{eqnarray}
w^t{\cal Z}\kappa \!\! &=& \!\! \frac{w^tC_{020}}\mu 
\nonumber \\
\!\! &=& \!\! -w^t \Omega'{\cal L}' + \rme^{-2\nu-2\psi}(w^t)^{-2}{\cal L}'{^2}
\nonumber \\
\!\! &=& \!\! -w^t \Omega'{\cal L}' -s^t {\cal E}' + s^\phi  {\cal L}' 
\nonumber \\
\!\! &=& \!\! -w^t \frac{{\cal L}'{\cal E}'' - {\cal E}'{\cal L}''}{{\cal L}'} -s^t {\cal E}' + s^\phi  {\cal L}' 
\nonumber \\
\!\! &=& \!\! -w^t{\cal E}'' + w^\phi{\cal L}'' -s^t {\cal E}' + s^\phi  {\cal L}' .
\label{eq:biglemma}
\end{eqnarray}
[Here the second line used Eq.~(\ref{eq:C020}); the third line used Eq.~(\ref{eq:slemma}); the fourth line used $\Omega={\cal E}'/{\cal L}'$ and the quotient rule; and the fifth line used that $w^\phi/w^t=\Omega={\cal E}'/{\cal L}'$.] Equation~(\ref{eq:biglemma}) leads to two major simplifications in Eq.~(\ref{eq:I2-intermed}). The term involving $\xi^r$ simplifies dramatically.  Also, using the first and third lines of Eq.~(\ref{eq:biglemma}) and ${\cal E}'=\Omega{\cal L}'$, we find that
\begin{equation}
-\Omega'w^t + s^\phi - \Omega_{\rm s}s^t = (\Omega-\Omega_{\rm s})s^t + w^t{\cal L}'{^{-1}}{\cal Z}\kappa.
\end{equation}
Therefore, Eq.~(\ref{eq:I2-intermed}) simplifies to
\begin{eqnarray}
I_2 &=& \!\!\!
\rmi m [ (\Omega_{\rm s}-\Omega)(w^t{_{,r}}-s^t) + w^t{\cal L}'{^{-1}}{\cal Z}\kappa]
(-{\cal E}\xi^t + {\cal L}\xi^\phi)
\nonumber \\ && \!\!\!
+ \rmi m(\Omega_{\rm s} - \Omega)w^t
(- {\cal E}\xi^t{_{,r}} + {\cal L}\xi^\phi{_{,r}})
\nonumber \\ && \!\!\!
+ w^t{\cal Z}\kappa \xi^r.
\label{eq:I2-intermed2}
\end{eqnarray}

A final level of simplification involves $w^t{_{,r}}-s^t$. Using the explicit expressions, Eq.~(\ref{eq:wt}) for $w^t$ and Eq.~(\ref{eq:s-component}) for $s^t$, we see that
\begin{eqnarray}
w^t{_{,r}} - s^t \!\! &=& \!\! \rme^{-2\nu} \left[ -2\nu_{,r} ({\cal E}-\oomega{\cal L}) + {\cal E}' - \oomega_{,r}{\cal L} - \oomega{\cal L}' \right]
\nonumber \\ 
&& \!\! - \rme^{-2\nu} ({\cal E}'-\oomega{\cal L}').
\nonumber \\
&=& - \left( 2\nu_{,r} - \frac{\oomega_{,r}{\cal L}}{{\cal E}-\oomega{\cal L}} \right) w^t.
\end{eqnarray}
This allows us to eliminate ${\bmath s}$ from our expression for $I_2$:
\begin{eqnarray}
I_2 \!\!\! &=& \!\!\!
\rmi m \Bigl[ -(\Omega_{\rm s}-\Omega)\left( 2\nu_{,r} - \frac{\oomega_{,r}{\cal L}}{{\cal E}-\oomega{\cal L}} \right) w^t
\nonumber \\ &&
 + w^t{\cal L}'{^{-1}}{\cal Z}\kappa\Bigr]
(-{\cal E}\xi^t + {\cal L}\xi^\phi)
\nonumber \\ && \!\!\!
+ \rmi m(\Omega_{\rm s} - \Omega)w^t
(- {\cal E}\xi^t{_{,r}} + {\cal L}\xi^\phi{_{,r}})
\nonumber \\ && \!\!\!
+ w^t{\cal Z}\kappa \xi^r.
\label{eq:I2}
\end{eqnarray}

\subsubsection{Evaluation of $I_3$}

Finally, we consider $I_3 = h^{r\alpha}w_\alpha$.  This is most easily computed by explicit evaluation of the contravariant components using Eq.~(\ref{eq:habinv}):
\begin{equation}
h^{rt} = -\rme^{-2\mmu}\xi^t{_{,r}} + \rmi m (\oomega-\Omega_{\rm s}) \rme^{-2\nu}\xi^r
\end{equation}
and
\begin{equation}
h^{r\phi} = -\rme^{-2\mmu}\xi^\phi{_{,r}} + \rmi m (\oomega-\Omega_{\rm s}) \rme^{-2\nu}\oomega\xi^r - \rmi m \rme^{-2\psi}\xi^r.
\end{equation}
This implies
\begin{eqnarray}
h^{r\alpha}w_\alpha &=& \rme^{-2\mmu}({\cal E}\xi^t{_{,r}}-{\cal L}\xi^\phi{_{,r}})
\nonumber \\ &&
  + \rmi m (\oomega-\Omega_{\rm s}) \rme^{-2\nu}\xi^r(-{\cal E}+\oomega{\cal L})
\nonumber \\ &&
  - \rmi m \rme^{-2\psi}{\cal L}\xi^r.
\end{eqnarray}

The terms involving $\xi^r$ can be simplified using Eqs.~(\ref{eq:wt}) and (\ref{eq:wphi}), which simplifies them to $\rmi m (\Omega_{\rm s}w^t-w^\phi)\xi^r$.  Further using $w^\phi=\Omega w^t$ gives
\begin{equation}
I_3=
h^{r\alpha}w_\alpha
= \rme^{-2\mmu}({\cal E}\xi^t{_{,r}}-{\cal L}\xi^\phi{_{,r}}) + \rmi m (\Omega_{\rm s}-\Omega)w^t\xi^r.
\end{equation}
The other contributions to ${\cal S}$ do not explicitly contain $\mmu$, so in order to prove gauge invariance we will need to eliminate $\mmu$ in favour of other variables.  Equation~(\ref{eq:C002}) provides a convenient choice: it and the definitions of $\kappa$ and ${\cal Z}$ tell us that
\begin{equation}
\rme^{-2\mmu} = \mu w^tC_{002} = w^t\frac\kappa{\cal Z}.
\end{equation}
We thus arrive at our final expression for $I_3$:
\begin{equation}
I_3= w^t\frac\kappa{\cal Z}({\cal E}\xi^t{_{,r}}-{\cal L}\xi^\phi{_{,r}}) + \rmi m (\Omega_{\rm s}-\Omega)w^t\xi^r.
\label{eq:I3}
\end{equation}

\subsubsection{Putting it all together}

We now substitute $I_1$, $I_2$, and $I_3$ into Eq.~(\ref{eq:FSM2}), giving
\begin{eqnarray}
w^t{\cal S} \!\!\! &=& \!\!\! \left[ -\frac m{{\cal L}'} \mp \frac1{\cal Z}\left(
2\nu_{,r} + \frac{\oomega_{,r}{\cal L}}{{\cal E}-\oomega{\cal L}}
\right)\right]
\rmi m(\Omega_{\rm s} - \Omega)
\nonumber \\ && \times
w^t(-{\cal E}\xi^t + {\cal L}\xi^\phi)
\nonumber \\ && \!\!\!
\mp \rmi \frac m{\cal Z} \Bigl[ -(\Omega_{\rm s}-\Omega)\left( 2\nu_{,r} - \frac{\oomega_{,r}{\cal L}}{{\cal E}-\oomega{\cal L}} \right) w^t
\nonumber \\ &&
 + \frac{w^t{\cal Z}\kappa}{{\cal L}'} \Bigr]
(-{\cal E}\xi^t + {\cal L}\xi^\phi)
\nonumber \\ && \!\!\!
\mp \rmi \frac m{\cal Z} (\Omega_{\rm s} - \Omega)w^t
(- {\cal E}\xi^t{_{,r}} + {\cal L}\xi^\phi{_{,r}})
\mp w^t \kappa \xi^r
\nonumber \\ && \!\!\!
+\rmi w^t\frac\kappa{\cal Z}({\cal E}\xi^t{_{,r}}-{\cal L}\xi^\phi{_{,r}}) - m (\Omega_{\rm s}-\Omega)w^t\xi^r.
\end{eqnarray}
We may divide through by $w^t$ on both sides, and cancel the terms involving $2\nu_{,r}- \oomega_{,r}{\cal L}/({\cal E}-\oomega{\cal L})$. Collecting the remaining terms gives
\begin{eqnarray}
{\cal S} \!\!\! &=& \!\!\! \rmi \frac m{{\cal L}'} (-{\cal E}\xi^t + {\cal L}\xi^\phi) [ m(\Omega-\Omega_{\rm s}) \mp \kappa]
\nonumber \\ && \!\!\!
+ \rmi \frac{-\kappa  \pm m (\Omega - \Omega_{\rm s})}{\cal Z} 
(- {\cal E}\xi^t{_{,r}} + {\cal L}\xi^\phi{_{,r}})
\nonumber \\ && \!\!\!
+ [m (\Omega-\Omega_{\rm s}) \mp\kappa ]\xi^r.
\label{eq:SM-simple}
\end{eqnarray}

In general, this is nonzero.  However, there is one piece of information we have not used: that the resonant amplitude is to be evaluated at the resonance location $D(R)=0$, i.e.
\begin{equation}
m(\Omega-\Omega_{\rm s}) = \pm\kappa.
\end{equation}
When -- and only when -- we use this fact, we see that Eq.~(\ref{eq:SM-simple}) vanishes.  That is, {\em the resonant amplitude ${\cal S}^{(m)}$ is only gauge-invariant when evaluated at the resonant position}!  This is not a problem since the torque formula contains a $\delta$-function at the resonance.

Thus we see that a pure gauge perturbation leads to zero contribution to ${\cal S}^{(m)}$ at resonance, and the resonant torque is gauge-invariant.

\subsection{Epicyclic geodesic formulation}

There is an alternative way of writing the resonant amplitude ${\cal S}^{(m)}$ that will be better suited to computation in the Schwarzschild and Kerr spacetimes.  We will argue in this section that ${\cal S}^{(m)}$ is related to a particular integral of the metric perturbation along the world line of a test particle on an orbit with very small eccentricity.  This formulation has some utility in the Newtonian case, but it will be shown to be very powerful in Paper II, where we will relate it to the gravitational waveform emitted by a test particle on such an orbit.  It will thus allow computation of ${\cal S}^{(m)}$ using standard methods for computing waveforms, without explicit evaluation of the metric perturbations.

Our starting point is to consider a particle on an unperturbed orbit (i.e. traveling according to $H_0$) oscillating between $r=R-\epsilon$ and $R+\epsilon$.  To first order in $\epsilon$, its trajectory is given by\footnote{These equations may be obtained from the Green's function relations in Section~\ref{ss:Green} by taking a particle in a circular orbit at radius $R$ that passes longitude $\phi=0$ at $t=0$, applying a perturbation at time $t=0$ that increments $r$ by $\diamondsuit r=\epsilon$, and considering the solution at $t>0$.}
\begin{eqnarray}
r \!\! &=& \!\! R + \epsilon \cos (\kappa t),
\nonumber \\
p_r \!\! &=& \!\! \mass \epsilon{\cal Z}\sin (\kappa t),
\nonumber \\
\phi \!\! &=& \!\! \Omega t + \epsilon \frac{C_{110}}\kappa \sin(\kappa t),
{\rm ~~and} \nonumber \\
p_\phi \!\! &=& \!\! \mass{\cal L}.
\end{eqnarray}

Now we consider the integral of the metric perturbation over the test particle world line
\begin{eqnarray}
\epsilon I_{\rm T}
\!\! &=& \!\! \int_{t_1}^{t_1+2\pi/\kappa} h_{\alpha\beta} \frac{u^\alpha u^\beta}{u^t} \,\rmd t
\nonumber \\
&=& \!\! \mass^{-1}\int_{\cal V} h^{(m)}_{\alpha\beta} \tilde T^{\alpha\beta} \,\sqrt{-\det g}\,\rmd^4{\bmath x},
\label{eq:IT}
\end{eqnarray}
where the range of integration is over any epicyclic period, i.e. from $t_1<t<t_1+2\pi/\kappa$ for any $t_1$; and in the second integral, $\tilde T^{\alpha\beta}$ is the stress-energy tensor associated with the test particle and ${\cal V}$ is the region of 4-volume in this range of coordinate time.  By construction, $I_{\rm T}$ is linear in the metric perturbation. It is also invariant under gauge transformations respecting the helical symmetry $\xi_\alpha^{(m)}\propto \rme^{\rmi m(\phi-\Omega_{\rm s}t)}$, since under a gauge transformation, Eq.~(\ref{eq:IT}) changes by
\begin{equation}
\epsilon \Delta I_{\rm T} = -\frac2\mass \int_{\cal V} \xi^{(m)}_{\alpha;\beta} \tilde T^{\alpha\beta} \,\sqrt{-\det g}\,\rmd^4{\bmath x}.
\end{equation}
We may integrate by parts to move the $_{;\beta}$ derivative onto $\tilde T^{\alpha\beta}$; but $\tilde T^{\alpha\beta}{_{;\beta}}=0$ for a test particle traveling along a geodesic.  The boundary terms at $t=t_1$ and $t_1+2\pi/\kappa$ also cancel each other since both $\xi^{(m)}_\alpha$ and $\tilde T^{\alpha\beta}$ are invariant under translation in time and longitude by $t\rightarrow t+2\pi/\kappa$ and
\begin{equation}
\phi\rightarrow\phi+\Omega\frac{2\pi}\kappa = \phi+\Omega_{\rm s}\frac{2\pi}\kappa \mp \frac{2\pi}m,
\end{equation}
respectively. Thus $\Delta I_{\rm T}=0$, and $I_{\rm T}$ is gauge-invariant.

We may use the gauge invariance of $I_{\rm T}$ and ${\cal S}^{(m)}$: if a relation between them can be demonstrated in one gauge, then it must be valid in any gauge.  We choose the gauge with $h^{tt}=h^{t\phi}=h^{\phi\phi}=0$.  This gauge exists for the generic case where $m\neq0$, since one may write the gauge transformation relations
\begin{equation}
\left( \begin{array}{c} \Delta h^{tt} \\ \Delta h^{t\phi} \\ \Delta h^{\phi\phi} \end{array} \right)
=
\left( \begin{array}{ccc}
g^{tt}{_{,r}} & -2c_\phi & 0 \\ g^{t\phi}{_{,r}} & c_t & -c_\phi \\ g^{\phi\phi}{_{,r}} & 0 & 2c_t
\end{array} \right)
\left( \begin{array}{c} \xi^r \\ \rmi m\xi^t \\ \rmi m\xi^\phi \end{array} \right),
\end{equation}
where we define the vector ${\bmath c}$ by $c_t = \Omega_{\rm s}g^{t\phi}-g^{\phi\phi}$, $c_\phi = g^{t\phi}-\Omega_{\rm s}g^{tt}$, and $c_r=c_z=0$.  If this $3\times 3$ matrix ${\mathbfss A}$ is nonsingular (which may be easily verified for some cases such as Schwarzschild), then the gauge $h^{tt}=h^{t\phi}=h^{\phi\phi}=0$ exists everywhere.  (We will remove the condition on ${\mathbfss A}$ later.)

In this gauge, we find from Eq.~(\ref{eq:FSM})
\begin{equation}
{\cal S}^{(m)} = \rmi \frac{ h^{(m)r\alpha}w_\alpha}{w^t}.
\end{equation}
We also find that in computing $I_{\rm T}$ only $h_{rt}$ and $h_{r\phi}$ contribute, and since $u^r$ is already ${\cal O}(\epsilon)$ we find
$u^r = \rme^{-2\mmu}u_r$ and
\begin{equation}
I_{\rm T} = 2\rme^{-2\mmu} {\cal Z}\int_{t_1}^{t_1+2\pi/\kappa}
\sin(\kappa t) \frac{h_{r\alpha}w^\alpha}{w^t} \rmd t.
\end{equation}
We may raise $r$ in the perturbation using the factor of $\rme^{-2\mmu}$, and simplify this to
\begin{equation}
I_{\rm T} = -2\rmi {\cal Z} {\cal S}^{(m)}|_0 \int_{t_1}^{t_1+2\pi/\kappa}
\sin(\kappa t)  \rme^{\rmi m(\phi-\Omega_{\rm s}t)} \rmd t,
\end{equation}
where the $|_0$ reminds us to evaluate ${\cal S}^{(m)}$ at $\phi=t=0$.  On resonance the complex exponential decomposition of the sine allows us to evaluate the integral to $-\rmi \pi/\kappa$, so we conclude that
\begin{equation}
I_{\rm T} = -2\pi \frac{\cal Z}\kappa {\cal S}^{(m)}.
\label{eq:ITSM}
\end{equation}
So long as $\det{\mathbfss A}\neq 0$,
this relation must be valid in all gauges since both sides are gauge-invariant.

In some spacetimes there are radii where $\det {\mathbfss A}=0$; however we may show Eq.~(\ref{eq:ITSM}) to be valid there as well.  We may consider a family of spacetimes ${\cal M}(P)$ whose metric tensor components are analytic in the parameter $P$, the desired spacetime is ${\cal M}(0)$, and $\det {\mathbfss A}\propto P^n$ ($n=1$, 2, or 3) for small $P$.  Then we may carry through the argument for slightly different values of the parameters controlling the spacetime and prove Eq.~(\ref{eq:ITSM}); then since both ${\cal S}^{(m)}$ and $I_{\rm T}$ are analytic and equal in a neighborhood of $P=0$ they must be equal at $P=0$.  Thus Eq.~(\ref{eq:ITSM}) remains valid regardless of whether $\det{\mathbfss A}=0$ or not.

We may then express ${\cal S}^{(m)}$ in terms of the integral of the metric perturbation against the stress-energy tensor of a test particle on a slightly eccentric orbit,
\begin{equation}
{\cal S}^{(m)} =
\frac{-\kappa}{2\pi\epsilon\mass{\cal Z}}\int_{\cal V} h^{(m)}_{\alpha\beta} \tilde T^{\alpha\beta} \,\sqrt{-\det g}\,\rmd^4{\bmath x}.
\end{equation}
A further simplification occurs if we extract the Fourier mode of frequency $-m\Omega_{\rm s}$ from the stress-energy tensor, 
which is the only one that can lead to a nonzero integral against $h^{(m)}$.  The $t$-integral is then trivial and we find
\begin{equation}
{\cal S}^{(m)} =
\frac{-1}{\epsilon\mass{\cal Z}}\int h^{(m)}_{\alpha\beta} \tilde T^{[-m\Omega_{\rm s}]\alpha\beta} \,\sqrt{-\det g}\,\rmd^3{\bmath x}.
\label{eq:SMIT}
\end{equation}

A second version of the epicyclic formulation is as follows.  We note that the average amount of power ${\cal P}$ transferred to the test particle by the perturbation in one radial cycle is
\begin{equation}
{\cal P} = \frac\kappa{2\pi}\int_{t_1}^{t_1+2\pi/\kappa} \dot H_1\,\rmd t;
\end{equation}
the integrand can be evaluated along the unperturbed trajectory since $H_1$ is already first-order, and we note that the dot pulls down a factor of $-\rmi m\Omega_{\rm s}$:
\begin{equation}
{\cal P} = \frac{\rmi m\Omega_{\rm s}\mu_1\kappa}{2\pi}\int_{t_1}^{t_1+2\pi/\kappa} h^{(m)\alpha\beta} \frac{u_\alpha u_\beta}{2u^t}\,\rmd t;
\end{equation}
the integral simplifies to $\frac12\epsilon I_{\rm T}$, and so we conclude that
\begin{equation}
{\cal P} = \frac{\rmi m\Omega_{\rm s}\mu\kappa}{4\pi} \epsilon I_{\rm T}.
\end{equation}
From this, we extract a relation between ${\cal S}^{(m)}$ and the power provided to a particle on a slightly eccentric orbit:
\begin{equation}
{\cal S}^{(m)} = \frac{2\rmi}{m\Omega_{\rm s}\mu{\cal Z}\epsilon} {\cal P}.
\label{eq:SP}
\end{equation}
Note that ${\cal P}$ pertains to the particlar $m$-mode and hence may be complex if the peak power occurs at a resonant phase other than $0$ or $\pi$.

\section{Newtonian Keplerian limit}
\label{sec:K}

We now consider the limit of Eq.~(\ref{eq:TX1}) for non-relativistic Newtonian-Keplerian disks, and show that it reduces to the familiar result.

In the limit of $M/r\ll 1$ where we expect to recover the Newtonian result, the metric for a central object of mass $M$ has $\nu= -Mr^{-1} \ll 1$, $\psi=\ln r$, and $\oomega=\mmu=0$.  The Hamiltonian evaluated at zero radial momentum, Eq.~(\ref{eq:X2}), is
\begin{eqnarray}
H(p_\phi,r,0) &=& \rme^{-M/r} \sqrt{\mass + (p_\phi/r)^2} 
\nonumber \\ &\approx& \mass\left[ 1 - \frac Mr + \frac{(p_\phi)^2}{2\mass^2r^2} \right];
\label{eq:Hcirc}
\end{eqnarray}
minimizing over $r$ gives $r=(p_\phi)^2/(M\mass^2)={\cal L}^2/M$, so
\begin{equation}
{\cal L} = M^{1/2}r^{1/2}.
\end{equation}
The energy is obtained by substituting back into Eq.~(\ref{eq:Hcirc}),
\begin{equation}
{\cal E} = 1 - \frac M{2r}.
\end{equation}
Using the results from Section~\ref{ss:relations}, we find
\begin{equation}
\Omega = M^{1/2}r^{-3/2}.
\end{equation}
Then $w^t = \rme^{-2\nu}({\cal E}-\oomega{\cal L}) = 1+\frac32Mr^{-1}$, so we find to leading order
\begin{equation}
C_{002} = \frac1{\mass} {\rm ~~and~~} C_{020} = \frac {M\mass}{r^3}.
\end{equation}
This implies the epicyclic frequency and specific impedance
\begin{equation}
\kappa = M^{1/2}r^{-3/2} {\rm ~~and~~} {\cal Z} = M^{1/2}r^{-3/2}.
\end{equation}

Since $\kappa=\Omega$ in the Newtonian Keplerian case, the Lindblad resonance condition $m(\Omega-\Omega_{\rm s})=\pm\kappa$ is satisfied for
\begin{equation}
\Omega = \frac m{m\mp1}\Omega_{\rm s} {\rm ~~or~~}
r = \left( \frac{m\mp 1}m \right)^{2/3}r_{\rm s}.
\end{equation}
We label the resonances with positive $m$ so that the lower sign corresponds to the OLRs and the upper sign to the ILRs.  In the Newtonian Keplerian case, there exist OLRs for each positive integer $m$, while the ILRs exist only for $m\ge 2$.

The resonant torque further involves the detuning function $dD/dr$:
\begin{equation}
D' = m\Omega' \mp \kappa' = (m\mp 1)\Omega' = -\frac32(m\mp 1)M^{1/2}r^{-5/2}.
\label{eq:Dprime}
\end{equation}

We now consider the resonant amplitude ${\cal S}^{(m)}$.  In the nonrelativistic limit, the time-time metric coefficient is
\begin{equation}
h^{tt} = 2\Phi,
\end{equation}
where $\Phi$ is the Newtonian gravitational potential associated with the perturbation, and the other components are small.\footnote{For $r/r_{\rm s}$ of order unity, ${\cal L}/{\cal E}\sim (r_{\rm s}/M)^{1/2}$, but the components such as $h^{r\phi}$ for nonrelativistic perturbers are suppressed by higher powers of $M/r_{\rm s}$.}  Keeping the leading-order ($r^{1/2}h^{tt}$) terms in Eq.~(\ref{eq:SM-explicit}) gives
\begin{equation}
{\cal S}^{(m)} = M^{-1/2}[-2mr^{1/2}\Phi^{(m)} \mp r^{3/2}\Phi^{(m)}{_{,r}}].
\end{equation}

The Keplerian analogue of the binary black hole case is for the perturber to be a point particle with mass $qM$, where the mass ratio $q\ll 1$.  Without loss of generality we may place the perturber at longitude $\lambda=0$; any other choice of longitude would result in $\Phi^{(m)}$ and hence ${\cal S}^{(m)}$ being multiplied by a factor of $\rme^{-\rmi m\lambda}$, which will have no effect on $|{\cal S}^{(m)}|^2$ or on the torque formula.

If we place this particle at radius $r_{\rm s}$, its perturbing potential is
\begin{equation}
\Phi(r,\phi) = qM\left[ -\frac 1{\sqrt{r_{\rm s}^2+r^2-2r_{\rm s}r\cos\phi}} + \frac 1{r_{\rm s}^2}r\cos\phi \right],
\label{eq:PhiNewt}
\end{equation}
where the second term is the ``indirect'' term resulting from the acceleration of the primary (i.e. it is necessary to keep the primary at the centre of the coordinate system, as is standard practice in celestial mechanics).
We may project out the order $m$ Fourier coefficient,
\begin{eqnarray}
\Phi^{(m)}(r,0) &=& \int_0^{2\pi} \frac{\rmd\phi}{2\pi} \rme^{-\rmi m\phi} \,\Phi(r,\phi)
\nonumber \\
&=& \frac {qM}{2 r_{\rm s}}[ -b_{1/2}^{(m)}(\varsigma) + \varsigma\delta_{m1} ],
\end{eqnarray}
where $\varsigma\equiv r/r_{\rm s}=[(m\mp 1)/m]^{2/3}$ and $b$ represents a Laplace coefficient (e.g. Eq.~6.67 of \citealt{2000ssd..book.....M}).  Then ${\cal S}^{(m)}$ (evaluated at zero longitude) is
\begin{eqnarray}
{\cal S}^{(m)} &=& \frac {qM^{1/2}}{2r_{\rm s}^{1/2}} \Bigl[ 2m\varsigma^{1/2}b_{1/2}^{(m)}(\varsigma) - 2\varsigma^{3/2}\delta_{m1}
\nonumber \\ &&
  \pm \varsigma^{3/2}b_{1/2}^{(m)}{'}(\varsigma) \mp \varsigma^{3/2}\delta_{m1}
\Bigr].
\end{eqnarray}
The $\delta_{m1}$ term exists only for the $m=1$ OLR (lower sign), so we may simplify this to
\begin{equation}
{\cal S}^{(m)} = \frac {qM^{1/2}\varsigma^{1/2}}{2r_{\rm s}^{1/2}} \left[ 2mb_{1/2}^{(m)}(\varsigma) - \varsigma\delta_{m1}
  \pm \varsigma b_{1/2}^{(m)}{'}(\varsigma) \right].
\label{eq:SMK}
\end{equation}
\cmnt{
[As an aside, we note that the indirect term only contributed to the $m=1$ OLR at $\varsigma = 2^{2/3}$.  An alternative formulation of the problem would have worked in the inertial frame, in which the second term in Eq.~(\ref{eq:PhiNewt}) would have been replaced with the term $qMr_{\rm s}r^{-2}\cos\phi$ associated with the displacement of the primary.  Following this through to Eq.~(\ref{eq:SMK}) would replace the indirect term $-\varsigma\delta_{m1}$ with $-4\varsigma^{-2}\delta_{m1}$.  These are equal, but only at the resonance location $\varsigma=2^{2/3}$.  This is a manifestation of the gauge invariance of ${\cal S}^{(m)}$ at and only at the resonance locations.]
}

Substitution into Eq.~(\ref{eq:TX1}) then gives
\begin{eqnarray}
T &=& \mp\frac{2\pi}{3} \frac m{m\mp1} \mass r
\frac {q^2M\alpha}{4r_{\rm s}}
\delta(r-R_{\rm r})
\nonumber \\ && \times
\left|
2mb_{1/2}^{(m)}(\varsigma) - \varsigma\delta_{m1}
  \pm \varsigma b_{1/2}^{(m)}{'}(\varsigma)
\right|^2 .
\end{eqnarray}
Using $m/(m\mp1)=\varsigma^{-3/2}$, we reduce this to
\begin{eqnarray}
T &=& \mp \frac\pi6 \mass q^2M\varsigma^{1/2}\delta(r-R_{\rm r})
\nonumber \\ && \times
\left|
2mb_{1/2}^{(m)}(\varsigma) - \varsigma\delta_{m1}
  \pm \varsigma b_{1/2}^{(m)}{'}(\varsigma)
  \right|^2.
\end{eqnarray}
In terms of the disk surface density, $\Sigma = \mass\delta(r-R_{\rm r})/(2\pi r)$, this becomes
\begin{eqnarray}
T &=& \mp \frac{\pi^2 q^2\alpha^{1/2}}{3}rM\Sigma
\nonumber \\ && \times
\left|
2mb_{1/2}^{(m)}(\varsigma) - \varsigma\delta_{m1}
  \pm \varsigma b_{1/2}^{(m)}{'}(\varsigma)
  \right|^2.
\label{eq:Kepler}
\end{eqnarray}

At $m\gg 1$ or $|\varsigma-1|\ll 1$ we may meaningfully consider the smoothed torque density over many resonances.\footnote{Whether $\rmd T/\rmd r$ is really smooth depends on the nature of the dissipation mechanism, which we do not consider here.}  Noting that $m(1-\varsigma)=\pm\frac23$, we find using the large $m$ expansion of the Laplace coefficient \citep{1980ApJ...241..425G}:
\begin{equation}
b_{1/2}^{(m)}(\varsigma) \approx \frac2\pi K_0(\smfrac23) {\rm~~and~~}
 b_{1/2}^{(m)}{'}(\varsigma)\approx \pm\frac{2m}\pi K_1(\smfrac23).
\end{equation}
Then for large $m$, Eq.~(\ref{eq:Kepler}) becomes
\begin{equation}
T = \mp\frac43m^2q^2rM\Sigma \left[ 2K_0(\smfrac23) + K_1(\smfrac23) \right]^2.
\end{equation} 
This is for a single resonance.  For a continuum of resonances, we need to substitute the resonance order $m=2r_{\rm s}/(3|r_{\rm s}-r|)$ and multiply by the density of resoances $|\rmd m/\rmd r|$
to get the torque density
\begin{equation}
\frac{\rmd T}{\rmd r} = \mp\frac{32}{81} \frac{q^2r_{\rm s}^4M\Sigma}{(r_{\rm s}-r)^4} \left[ 2K_0(\smfrac23) + K_1(\smfrac23) \right]^2,
\end{equation}
which agrees with Eq.~(18) of \citet{1980ApJ...241..425G}.

\section{Relativistic disc heating and surface brightness}
\label{sec:heat}

Thus far, we have considered the angular momentum and energy transfer to the disk at the Lindblad resonances.  In Newtonian thin-disk problems, it is often the case that the disc can radiate energy but not angular momentum.  Since an orbit of fixed angular momentum has a minimum possible energy, one can then compute the rate of energy input that does {\em not} go into orbital energy; this amount of energy goes into epicyclic motions, which are eventually converted to heat and ultimately radiated.  The relativistic case is far more complicated because radiation carries away both energy {\em and} angular momentum.  We shall consider the problem here under the following two simplifying assumptions:
\begin{itemize}
\item The dissipative process is localized, i.e. the energy of epicyclic motions is dissipated near the resonant radius rather than being transmitted to a distant part of the disc (e.g. via density waves).
\item The energy is radiated away locally, i.e. we assume a thin disk rather than an ADAF or other radiatively inefficient solution.
\end{itemize}

The second assumption is necessary in order to maintain a thin disc, i.e. for the consistency of this paper.  In some cases, it may well break down.  For example, in the problem of \citet{2009arXiv0906.0825C}, in which the secondary ``shepherds'' the inner disc to smaller radii, it is conceivable that heating from resonant torques could destroy the thin disc solution.  Even in this case, however, we would like to know the resonant heating formula for a thin disc: inability to produce the required flux $F$ for any disc temperature would be a sufficient condition for the destruction of the thin disc.

\subsection{Definitions and mathematical relations}

We use the formalism of \citet{1974ApJ...191..499P} to investigate the flux emerging from the disk, although we do {\em not} make the assumption that the disc is time-steady.  We {\em do} assume that the disc is thin and that the internal energy is negligible compared to the orbital energy, i.e. if the bulk 4-velocity of the baryonic material is ${\bmath u}$,
\begin{equation}
T^{\mu\nu} = \rho_0 u^\mu u^\nu + t^{\mu\nu} + u^\mu q^\nu + q^\mu u^\nu,
\end{equation}
where $\rho_0$ is the rest mass density (i.e. the mass of a baryon times the number density), ${\bmath q}$ is the heat flux, and $t^{\mu\nu}$ is the stress tensor in the baryon rest frame (by definition $q^\mu u_\mu = 0$ and $t^{\mu\nu} u_\mu=0$).  In accordance with \citet{1974ApJ...191..499P}, we assume that ${\bmath q}$ lies in the $z$-direction (i.e. $q^z$ is the only nonzero component).  The disc is assumed to be contained within a vertical thickness of $|z|<H$; the stress tensor at $z=\pm H$ is assumed to satisfy
\begin{equation}
t_\phi{^z} = t_r{^z} = t_t{^z} = 0.
\end{equation}
\citet{1974ApJ...191..499P} explicitly write time and longitude averages of these quantities, with the idea being to treat e.g. turbulent stresses as part of $t_\mu{^\nu}$ rather than as small-scale structure in ${\bmath u}$.  We will not write these averages explicitly, but note that (i) they are implied, and (ii) in our case, the time averaging is assumed to be over a duration long compared with the turnover time of turbulent eddies but short compared to the evolution timescales of the system (e.g. the merger timescale).  It is assumed that the disc material is on nearly circular orbits, but possibly with a small radial velocity, i.e. $u^t=w^t$, $u^\phi=w^\phi$, $u^z=0$, and $|u^r|\ll \sqrt{M/r}$.

We define the integrated quantities through the disc: the surface density,
\begin{equation}
\Sigma (r,t) \equiv\int_{-H}^{H} \rho_0(r,t,z)\,\rmd z,
\end{equation}
and the integrated shear stress,
\begin{equation}
W_{\phi}{^r}(r,t) \equiv\int_{-H}^H t_\phi{^r}(r,t,z)\,\rmd z.
\end{equation}
We also define the one-sided emergent flux
\begin{equation}
F(r,t) \equiv q^z(r,t,z=H) = -q^z(r,t,z=-H),
\end{equation}
which is the flux that would be seen by an observer sitting at the disc photosphere and corotating with the disc \citep{1974ApJ...191..499P}.

We further neglect stresses in the tangential direction, i.e. we set $t_\phi{^\phi}=0$.  Orthogonality with ${\bmath u}$ then implies $t_\phi{^t}=0$.
We note that the requirement that $t_\mu{^\nu}u^\mu=0$, combined with the approximation that ${\bmath u}\approx{\bmath r}$, gives us the integral
\begin{equation}
\int_{-H}^H t_t{^r}(r,t,z)\,\rmd z = -\Omega W_{\phi}{^r}(r,t).
\end{equation}

\subsection{Conservation laws}

As is the case with the time-steady thin accretion disk, it is convenient to use the conservation of baryonic rest mass, angular momentum, and energy to solve for the state of the system.  In our case, the equations will be time-dependent but their derivation is similar.  For any current ${\bmath j}$ satisfying
\begin{equation}
j^\alpha{_{;\alpha}} = \Gamma,
\end{equation}
where the source term $\Gamma$ is the amount of charge added per unit proper 4-volume,
we have\footnote{We have used the general expression for the divergence, $j^\alpha{_{;\alpha}}=(-|g|)^{-1/2}\partial_\alpha[(-|g|)^{1/2}j^\alpha]$, and recalled that for our choice of coordinates $-|g|^{1/2}=r$.}
\begin{equation}
0 = \frac{\partial}{\partial x^\alpha}(rj^\alpha) + r\Gamma;
\end{equation}
integrating this equation over $z$ from $-H$ to $H$, we get:
\begin{eqnarray}
0 &=&\frac{\partial}{\partial r}\left(r \int_{-H}^H j^r\,\rmd z\right)
 + \frac{\partial}{\partial t}\left(r \int_{-H}^H j^t\,\rmd z\right)
\nonumber \\ &&
 + r[j^z(z=H)-j^z(z=-H)] + r\int_{-H}^H \Gamma\,\rmd z.
\end{eqnarray}
This implies, for the rest-mass current $^{\rm(m)}j^\mu=\rho_0u^\mu$, which has no source,
\begin{equation}
0 = \frac{\partial}{\partial r}(r\Sigma u^r) + rw^t\dot\Sigma.
\label{eq:cM}
\end{equation}
For the angular momentum current $^{\rm(L)}j^\mu = T_\phi{^\mu}$, there is a source, namely there is an angular momentum $\rmd T/\rmd r$ added per unit radial coordinate per unit coordinate time.  This is related to the source via
\begin{equation}
\frac{\rmd T}{\rmd r} = 2\pi r \int_{-H}^H {^{\rm(L)}}\Gamma\,\rmd z.
\end{equation}
Thus
\begin{equation}
0 = \frac{\partial}{\partial r}(rW_\phi{^r} + r{\cal L}\Sigma u^r)
  + r{\cal L}w^t\dot\Sigma
  + 2r{\cal L}F + \frac1{2\pi}\frac{\rmd T}{\rmd r}.
\label{eq:cL}
\end{equation}
For the energy current $^{\rm(E)}j^\mu = -T_t{^\mu}$, the source differs from the angular momentum source in that the energy added is equal to $\Omega_{\rm s}$ times the angular momentum added.  This is a direct consequence of the fact that the time dependence of the metric perturbation consists solely of a pattern speed $\Omega_{\rm s}$.
Then $^{\rm(E)}\Gamma = \Omega_{\rm s}{^{\rm(L)}}\Gamma$, so
\begin{equation}
0 = \frac{\partial}{\partial r}(r\Omega W_\phi{^r} + r{\cal E}\Sigma u^r)
 + r{\cal E}w^t\dot\Sigma + 2r{\cal E}F + \frac{\Omega_{\rm s}}{2\pi}\frac{\rmd T}{\rmd r}.
\label{eq:cE}
\end{equation}

Equations~(\ref{eq:cM}), (\ref{eq:cL}), and (\ref{eq:cM}) provide 3 constraints for 4 unknowns ($\dot\Sigma$, $u^r$, $F$, and $W_\phi{^r}$).  They can be solved if a prescription is available for the shear stress $W_\phi{^r}$, e.g. an $\alpha$-prescription \citep{1973A&A....24..337S}.\footnote{\citet{1974ApJ...191..499P} were able to solve this system in the time-steady case without assuming any prescription for angular momentum transport by setting $\dot\Sigma\rightarrow 0$ and using the 3 equations to solve for the remaining unknowns $u^r$, $F$, and $W_\phi{^r}$.  This method is clearly not applicable to a transient event such as an inspiral.}

\subsection{No-viscosity solution}

A special case of interest to us is the case where the viscosity of the disc is negligible ($W_\phi{^r}=0$).  This limit is appropriate in the final stages of a binary black hole inspiral where the viscous timescale becomes short compared to the merger timescale, as occurs in the \citet{2009arXiv0906.0825C} calculation.  Then the disc evolution is dominated by angular momentum transport via the resonances and by the inspiral of the secondary black hole (itself driven by radiation reaction).

Writing Eq.~(\ref{eq:cL}) without the $W_\phi{^r}$ term, and using Eq.~(\ref{eq:cM}) to eliminate $\dot\Sigma$, we find
\begin{equation}
r{\cal L}' \Sigma u^r + 2r{\cal L}F = \frac1{2\pi}\frac{\rmd T}{\rmd r}.
\end{equation}
Similarly, using Eq.~(\ref{eq:cE}) gives
\begin{equation}
r{\cal E}' \Sigma u^r + 2r{\cal E}F = \frac{\Omega_{\rm s}}{2\pi}\frac{\rmd T}{\rmd r}.
\end{equation}
This gives us a linear system for $u^r$ and $F$, with solution:
\begin{equation}
u^r = \frac{1}{2\pi r\Sigma}\frac{\rmd T}{\rmd r}\,\frac{{\cal E}-\Omega_{\rm s}{\cal L}}{{\cal EL}'-{\cal LE}'}
\end{equation}
and
\begin{equation}
F = \frac{1}{4\pi r}\frac{\rmd T}{\rmd r}\,\frac{\Omega_{\rm s}{\cal L}'-{\cal E}'}{{\cal EL}'-{\cal LE}'}.
\end{equation}

To proceed further, we use Eq.~(\ref{eq:dTdr}) in the flux equation.  We then reduce this using the relations
\begin{equation}
{\cal EL}'-{\cal LE}' = {\cal L}'({\cal E}-\Omega{\cal L})
\end{equation}
and
\begin{equation}
\Omega_{\rm s}{\cal L}'-{\cal E}' = {\cal L}'(\Omega_{\rm s}-\Omega) = \mp\frac\kappa m,
\end{equation}
yielding finally
\begin{equation}
F = \frac{\pi\kappa}{2r|D'|}
\frac{w^t}{{\cal E}-\Omega{\cal L}}
{\cal Z}\Sigma|{\cal S}^{(m)}|^2\delta(r-R_{\rm r}).
\label{eq:flux-diss}
\end{equation}
Thus the emerging flux is, as expected, proportional to the surface density of material at resonance and localized at the resonance.  In reality, the $\delta$-function would be smeared out in a way that depends on the dissipation mechanism.  We note further that $r$ is not a proper radial coordinate: an observer sitting on the disc would measure a proper radial distance element $\rme^\mmu\,\rmd r$ instead of $\rmd r$.  That is, the emitted flux per unit length (units: erg$\,$s$^{-1}\,$cm$^{-1}$) along the circumference as measured locally by an observer on the disc would be
\begin{equation}
\int F\,\rmd r_{\rm proper} = \frac{\pi\kappa\rme^{\mmu-2\nu}}{2r|D'|}{\cal Z}\Sigma|{\cal S}^{(m)}|^2.
\label{eq:FluxReq}
\end{equation}

Equation~(\ref{eq:FluxReq}) gives the emitted flux required for the disc to remain thin.  It is of course emitted over some finite range of radii: there is a finite damping region for the density waves excited at each Lindblad resonance, and turbulent diffusion may transfer some heat to neighboring parts of the disc.  If this amount of flux cannot be radiated by any viable disc model regardless of the temperature then the thin disc solution must fail.

\cmnt{
The solution for $u^r$ simplifies to
\begin{equation}
u^r = \mp \frac{\pi M}{r|D'|} \frac{w^t}{{\cal L}'} \frac{{\cal E}-\Omega_{\rm s}{\cal L}}{ {\cal E}-\Omega{\cal L}}{\cal Z}|{\cal S}^{}|^2\delta(r-R_{\rm r}).
\label{eq:ur-diss}
\end{equation}
}

\section{Summary}
\label{sec:disc}

This paper has worked out the general formula for the torque on an equatorial disc in a stationary, axisymmetric spacetime with an equatorial plane of symmetry due to Lindblad resonances associated with a perturbation.  We have shown that the torque formula is gauge-invariant, and that the familiar formula is recovered for the problem of a Newtonian Keplerian disc with a perturber on a circular orbit.  We have also obtained the expression for the radiated flux required to maintain a thin disc solution.

The most important astrophysical application of the relativistic torque formula is to the Schwarzschild and Kerr spacetimes.  The computation of the resonance locations and amplitudes for these cases is presented in the companion paper, Paper II.

\section*{Acknowledgments}

C.H. thanks Tanja Hinderer, Mike Kesden, and Dave Tsang for numerous helpful conversations.

C.H. is supported by the US National Science Foundation (AST-0807337), the US Department of Energy (DE-FG03-02-ER40701), and the Alfred P. Sloan Foundation.

\label{lastpage} 


\begin{thebibliography}{}

\bibitem[\protect\citeauthoryear{{Abramowitz} \& {Stegun}}{1972}]{1972hmfw.book.....A}
{Abramowitz} M., {Stegun} I., Handbook of Mathematical Functions, Dover, New York, NY

\bibitem[\protect\citeauthoryear{{Anderson} et~al.}{2010}]{2010PhRvD..81d4004A}
{Anderson} M., {Lehner} L., {Megevand} M., {Neilsen} D., 2010, Phys. Rev. D, 81, 044004

\bibitem[\protect\citeauthoryear{{Armitage} \& {Natarajan}}{2002}]{2002ApJ...567L...9A}
{Armitage} P., {Natarajan} P., 2002, ApJ, 567, L9

\bibitem[\protect\citeauthoryear{{Bode} \& {Phinney}}{2007}]{2007APS..APR.S1010B}
{Bode} N., {Phinney} E., 2007, American Physical Society Meeting, 2007 April 14--17, Abstract \#S1.010

\bibitem[\protect\citeauthoryear{{Chandrasekhar}}{1992}]{1992mtbh.book.....C}
{Chandrasekhar} S., 1992, The Mathematical Theory of Black Holes, Oxford University Press, New York, NY

\bibitem[\protect\citeauthoryear{{Chang} et~al.}{2010}]{2009arXiv0906.0825C}
{Chang} P., {Strubbe} L., {Menou} K., {Quataert} E., 2010, MNRAS, 407, 2007

\bibitem[\protect\citeauthoryear{{Goldreich} \& {Tremaine}}{1978}]{1978ApJ...222..850G}
{Goldreich} P., {Tremaine} S., 1978, ApJ, 222, 850

\bibitem[\protect\citeauthoryear{{Goldreich} \& {Tremaine}}{1979}]{1979ApJ...233..857G}
{Goldreich} P., {Tremaine} S., 1979, ApJ, 233, 857

\bibitem[\protect\citeauthoryear{{Goldreich} \& {Tremaine}}{1980}]{1980ApJ...241..425G}
{Goldreich} P., {Tremaine} S., 1980, ApJ, 241, 425



\bibitem[\protect\citeauthoryear{{Hughes}}{2000}]{2000PhRvD..61h4004H}
{Hughes} S., 2000, Phys. Rev. D, 61, 084004

\bibitem[\protect\citeauthoryear{{Infeld} \& {Pleba\'nski}}{1960}]{IP60}
{Infeld} L., {Pleba\'nski} J., 1960, Motion and Relativity, Pergamon Press, Oxford, UK

\bibitem[\protect\citeauthoryear{{Kocsis} \& {Loeb}}{2008}]{2008PhRvL.101d1101K}
{Kocsis} B., {Loeb} A., 2008, Phys. Rev. Lett., 101, 041101

\bibitem[\protect\citeauthoryear{{Lin} \& {Papaloizou}}{1979}]{1979MNRAS.186..799L}
{Lin} D., {Papaploizou} J., 1979, MNRAS, 186, 799

\bibitem[\protect\citeauthoryear{{Lubow} \& {Ogilvie}}{1998}]{1998ApJ...504..983L}
{Lubow} S., {Ogilvie} G., 1998, ApJ, 504, 983

\bibitem[\protect\citeauthoryear{{Lynden-Bell} \& {Kalnajs}}{1972}]{1972MNRAS.157....1L}
{Lynden-Bell} D., {Kalnajs} A., 1972, MNRAS, 157, 1

\bibitem[\protect\citeauthoryear{{MacFadyen} \& {Milosavljevi\'c}}{2008}]{2008ApJ...672...83M}
{MacFadyen} A., {Milosavljevi\'c} M., 2008, ApJ, 672, 83

\bibitem[\protect\citeauthoryear{{Meyer-Vernet} \& {Sicardy}}{1987}]{1987Icar...69..157M}
{Meyer-Vernet} N., {Sicardy} B., 1987, Icarus, 69, 157

\bibitem[\protect\citeauthoryear{{Milosavljevi\'c} \& {Phinney}}{2005}]{2005ApJ...622L..93M}
{Milosavljevi\'c} M., {Phinney} E., 2005, ApJ, 622, L93

\bibitem[\protect\citeauthoryear{{Murray} \& {Dermott}}{2000}]{2000ssd..book.....M}
{Murray} C., {Dermott} S., 2000, Solar System Dynamics, Cambridge University Press, Cambridge, UK

\bibitem[\protect\citeauthoryear{{Narayan}}{2000}]{2000ApJ...536..663N}
{Narayan} R., 2000, ApJ, 536, 663

\bibitem[\protect\citeauthoryear{{Ogilvie}}{2007}]{2007MNRAS.374..131O}
{Ogilvie} G., 2007, MNRAS, 374, 131

\bibitem[\protect\citeauthoryear{{Ohta} et~al.}{1973}]{1973PThPh..50..492O}
{Ohta} T., {Okamura} H., {Kimura} T., {Hiida} K., 1973, Prog. Theor. Phys., 50, 492

\bibitem[\protect\citeauthoryear{{Ortega-Rodr\'iguez} et~al.}{2002}]{2002ApJ...567.1043O}
{Ortega-Rodr\'iguez} M., {Silbergleit} A., {Wagoner} R., 2002, ApJ, 567, 1043

\bibitem[\protect\citeauthoryear{{Page} \& {Thorne}}{1974}]{1974ApJ...191..499P}
{Page} D., {Thorne} K., 1974, ApJ, 191, 499

\bibitem[\protect\citeauthoryear{{Perez} et~al.}{1997}]{1997ApJ...476..589P}
{Perez} C., {Silbergleit} A., {Wagoner} R., {Lehr} D., ApJ, 476, 589

\bibitem[\protect\citeauthoryear{{Rossi} et~al.}{2010}]{2010MNRAS.401.2021R}
{Rossi} E., {Lodato} G., {Armitage} P., {Pringle} J., {King} A., 2010, MNRAS, 401, 2021

\bibitem[\protect\citeauthoryear{{Schnittman}}{2010}]{2010arXiv1006.0182S}
{Schnittman} J., 2010, ApJ, submitted, preprint, arXiv:1006.0182

\bibitem[\protect\citeauthoryear{{Schnittman} \& {Krolik}}{2008}]{2008ApJ...684..835S}
{Schnittman} J., {Krolik} J., 2008, ApJ, 684, 835

\bibitem[\protect\citeauthoryear{{Shakura} \& {Sunyaev}}{1973}]{1973A&A....24..337S}
{Shakura} N., {Sunyaev} R., 1973, A\&A, 24, 337

\bibitem[\protect\citeauthoryear{{Shields} \& {Bonning}}{2008}]{2008ApJ...682..758S}
{Shields} G., {Bonning} E., 2008, ApJ, 682, 758

\bibitem[\protect\citeauthoryear{{Silbergleit} et~al.}{2001}]{2001ApJ...548..335S}
{Silbergleit} A., {Wagoner} R., {Ortega-Rodr\'iguez} M., ApJ, 548, 335


\bibitem[\protect\citeauthoryear{{Wald}}{1984}]{1984ucp..book.....W}
{Wald} R., 1984, General Relativity, University of Chicago Press, Chicago, IL

\end{thebibliography}
\end{document}